\documentclass[twocolumn, twocolappendix]{aastex631}

\usepackage{amsmath}
\usepackage{caption}
\usepackage{subcaption}
\usepackage{cleveref}
\usepackage{graphicx}

\begin{document}

\title{A large-scale ring galaxy at $z=2.2$ revealed by JWST/NIRCam: kinematic observations and analytical modelling}

\author[0000-0003-1785-1357]{A. Nestor Shachar}
\affiliation{School of Physics and Astronomy, Tel Aviv University, Tel Aviv 69978, Israel}

\author[0000-0001-5065-9530]{A. Sternberg}
\affiliation{School of Physics and Astronomy, Tel Aviv University, Tel Aviv 69978, Israel}
\affiliation{Max-Planck-Institut für Extraterrestrische Physik (MPE), Giessenbachstr.1, 85748 Garching, Germany}
\affiliation{Center for Computational Astrophysics, Flatiron Institute. 162 5th Avenue, New York, NY, 10010, USA}

\author[0000-0002-2767-9653]{R. Genzel}
\affiliation{Max-Planck-Institut für Extraterrestrische Physik (MPE), Giessenbachstr.1, 85748 Garching, Germany}

\author[0000-0001-9773-7479]{D. Liu}
\affiliation{Purple Mountain Observatory, Chinese Academy of Sciences, 10 Yuanhua Road, Nanjing 210023, China}

\author[0000-0002-0108-4176]{S. H. Price}
\affiliation{Department of Physics and Astronomy and PITT PACC, University of Pittsburgh, Pittsburgh, PA 15260, USA}

\author[0000-0002-1428-1558]{C. Pulsoni}
\affiliation{Max-Planck-Institut für Extraterrestrische Physik (MPE), Giessenbachstr.1, 85748 Garching, Germany}


\author[0000-0002-1485-9401]{L. J. Tacconi}
\affiliation{Max-Planck-Institut für Extraterrestrische Physik (MPE), Giessenbachstr.1, 85748 Garching, Germany}

\author[0000-0002-2775-0595]{R. Herrera-Camus}
\affiliation{Departamento de Astronomía, Universidad de Concepción, Barrio Universitario, Concepción, Chile}

\author[0000-0003-4264-3381]{N. M. F{\"o}rster Schreiber}
\affiliation{Max-Planck-Institut für Extraterrestrische Physik (MPE), Giessenbachstr.1, 85748 Garching, Germany}

\author[0000-0001-6879-9822]{A. Burkert}
\affiliation{Universitäts-Sternwarte Ludwig-Maximilians-Universität (USM), Scheinerstr. 1, München, D-81679, Germany}

\author[0000-0002-3405-5646]{J. B. Jolly}
\affiliation{Max-Planck-Institut für Extraterrestrische Physik (MPE), Giessenbachstr.1, 85748 Garching, Germany}

\author[0000-0003-0291-9582]{D. Lutz}
\affiliation{Max-Planck-Institut für Extraterrestrische Physik (MPE), Giessenbachstr.1, 85748 Garching, Germany}

\author[0000-0003-3735-1931]{S. Wuyts}
\affiliation{Department of Physics, University of Bath, Claverton Down, Bath, BA2 7AY, UK}

\author[0000-0002-1952-3966]{C. Barfety}
\affiliation{Max-Planck-Institut für Extraterrestrische Physik (MPE), Giessenbachstr.1, 85748 Garching, Germany}

\author[0000-0001-5301-1326]{Y. Cao}
\affiliation{Max-Planck-Institut für Extraterrestrische Physik (MPE), Giessenbachstr.1, 85748 Garching, Germany}

\author[0000-0003-3921-3313]{J. Chen}
\affiliation{Max-Planck-Institut für Extraterrestrische Physik (MPE), Giessenbachstr.1, 85748 Garching, Germany}

\author[0000-0003-4949-7217]{R. Davies}
\affiliation{Max-Planck-Institut für Extraterrestrische Physik (MPE), Giessenbachstr.1, 85748 Garching, Germany}

\author{F. Eisenhauer}
\affiliation{Max-Planck-Institut für Extraterrestrische Physik (MPE), Giessenbachstr.1, 85748 Garching, Germany}

\author[0000-0001-6703-4676]{J. M. Espejo Salcedo}
\affiliation{Max-Planck-Institut für Extraterrestrische Physik (MPE), Giessenbachstr.1, 85748 Garching, Germany}

\author[0000-0001-7457-4371]{L. L. Lee}
\affiliation{Max-Planck-Institut für Extraterrestrische Physik (MPE), Giessenbachstr.1, 85748 Garching, Germany}

\author[0000-0002-2419-3068]{M. Lee}
\affiliation{DTU-Space, Technical University of Denmark, Elektrovej 327, DK2800 Kgs. Lyngby, Denmark}
\affiliation{Cosmic Dawn Center (DAWN), Denmark}

\author[0000-0002-7314-2558]{T. Naab}
\affiliation{Max Planck Institute for Astrophysik (MPA), Karl-Schwarzschild-Str 1, D-85748 Garching, Germany}

\author[0009-0009-0472-6080]{S. Pastras}
\affiliation{Max-Planck-Institut für Extraterrestrische Physik (MPE), Giessenbachstr.1, 85748 Garching, Germany}
\affiliation{Max Planck Institute for Astrophysik (MPA), Karl-Schwarzschild-Str 1, D-85748 Garching, Germany}

\author[0000-0002-2125-4670]{T. T. Shimizu}
\affiliation{Max-Planck-Institut für Extraterrestrische Physik (MPE), Giessenbachstr.1, 85748 Garching, Germany}

\author[0000-0002-0018-3666]{E. Sturm}
\affiliation{Max-Planck-Institut für Extraterrestrische Physik (MPE), Giessenbachstr.1, 85748 Garching, Germany}

\author[0000-0003-4226-7777]{G. Tozzi}
\affiliation{Max-Planck-Institut für Extraterrestrische Physik (MPE), Giessenbachstr.1, 85748 Garching, Germany}

\author[0000-0003-4891-0794]{H. {\"U}bler}
\affiliation{Max-Planck-Institut für Extraterrestrische Physik (MPE), Giessenbachstr.1, 85748 Garching, Germany}

\begin{abstract}
A unique galaxy at $z=2.2$, zC406690, has a striking clumpy large-scale ring structure that persists from rest UV to near-infrared, yet has an ordered rotation and lies on the star-formation main sequence. We combine new \emph{JWST}/NIRCam and ALMA band 4 observations, together with previous VLT/SINFONI integral field spectroscopy and HST imaging to re-examine its nature. The high-resolution H$\alpha$ kinematics are best fitted if the mass is distributed within a ring with total mass $M_{\mathrm{ring}} \approx 2 \times 10^{10}\, \mathrm{M_\odot}$ and radius $R_{\mathrm{ring}} = 4.6\, \mathrm{kpc}$, together with a central undetected mass component (e.g., a ``bulge") with a dynamical mass of $M_{\mathrm{bulge}} = 8 \times 10^{10}\, \mathrm{M_{\odot}}$. We also consider a purely flux-emitting ring superposed over a faint exponential disk, or a highly ``cuspy" dark matter halo, both disfavored against a massive ring model. The low-resolution CO(4-3) line and 142GHz continuum emission imply a total molecular and dust gas masses of $M_{\mathrm{mol,gas}}=7.1 \times 10^{10} \, \mathrm{M_{\odot}}$ and $M_{\mathrm{dust}} = 3 \times 10^{8} \, \mathrm{M_{\odot}}$ over the entire galaxy, giving a dust-to-mass ratio of $0.7 \%$. We estimate that roughly half the gas and dust mass lie inside the ring, and that $\sim$10\% of the total dust is in a foreground screen that attenuates the stellar light of the bulge in the rest-UV to near-infrared. Sensitive high-resolution ALMA observations will be essential to confirm this scenario and study the gas and dust distribution.
\end{abstract}

\submitjournal{ApJ February 26, 2025. Accepted June 5, 2025.}



\section{Introduction}\label{sec:1.intro}
Massive star-forming galaxies (SFGs) at the peak of cosmic star-formation (SF) around $z \sim 2$ are known to be predominantly rotationally supported disks, and typically follow a tight relation between their star-formation rate (SFR) and stellar mass $M_\star$, known as the ``main sequence" (MS) \citep{Madau2014, Speagle2014, ForsterSchreiber2018, Wisnioski2019, Freundlich2019, ferreira_jwst_2023, vanderWel2024}. The high SFRs are driven by large reservoirs of cold gas, often comparable to the entire stellar mass of the galaxy \citep{Tacconi2020, ForsterSchreiber2020}. Morphologically, these galaxies are often populated with giant SF clumps which can be very bright in UV, leading to irregular morphologies when observed at these rest-frame wavelengths, as is the case with \emph{HST} at $z=1-2$ \citep{genzel_sins_2011, Swinbank2012, ForsterSchreiber2018, ferreira_jwst_2023}. These clumps are believed to be formed through gravitational fragmentation via Toomre instabilities, driven by high gas fractions and large turbulent motions \citep{Toomre1964,Ubler2019}. However, in rest-frame near-infrared the light distribution is smoother and more symmetric, following more closely the stellar mass distribution \citep[e.g.,][]{ferreira_jwst_2023, Kartaltepe2023}. 

Detailed kinematics have become an important tool accessible to study individual galaxies at high redshift \citep{Herrera-Camus2022, Rizzo2023, Genzel2023, nestor_shachar_rc100_2023, Ubler2024_GANIFS, Roman-Oliveira2024, Birkin2024}. Ground-based Integral-Field Spectroscopy has been invaluable for observing the ionized gas phase of galaxies, mostly through H$\alpha$ emission which is redshifted to the near-IR at $z \sim 1-3$ (such as with KMOS$^{\mathrm{3D}}$; \citealt{Wisnioski2015}; SINS/zC-SINF, \citealt{ForsterSchreiber2009, ForsterSchreiber2018}; MUSE Extremely Deep Field (MXDF), \citealt{Bouche2022}; KMOS Redshift One Spectroscopic Survey (KROSS), \citealt{Stott2016}; and the ongoing GALPHYS program with the high-resolution spectrograph VLT/ERIS, (Förster Schreiber et al., in prep.)). In addition, sub-mm interferometry traces the cold gas phase, through, for example, the $\mathrm{^{12}CO}$ rotational $J \rightarrow J-1$ emission lines (such as the PHIBSS survey, \citealt{Tacconi2013, Freundlich2019}; the ongoing \emph{IRAM}/NOEMA$^{\mathrm{3D}}$ large program; ALMA-ALPAKA survey \citealt{Rizzo2023} and ALMA-CRISTAL survey \citealt{Telikova2024}). Kinematic modeling provides the dynamical mass content of the galaxy, its rotational stability, and dark-to-baryonic mass ratios \citep[e.g.][]{Genzel2017, Price2021_rc41, Bouche2022}. Such studies find that at $z \approx 2$ massive SFGs are frequently baryon dominated within the disk effective radius, quantified by the Dark Matter (DM) mass fraction within the effective radius $f_{\mathrm{DM}}(R_{\mathrm{eff}}) = V_{\mathrm{DM}}^2(R_{\mathrm{eff}}) / V_{\mathrm{circ}}^2(R_{\mathrm{eff}})$, with median values of $f_{\mathrm{DM}}(R_{\mathrm{eff}}) = 0.27$ \citep{Lang2017, Genzel2017, Genzel2020, Price2021_rc41, nestor_shachar_rc100_2023}.

One of these previously studied galaxies is zC406690, a massive MS SFG at $z=2.196$ with a total stellar mass from SED fitting of $M_\star = 4 \times 10^{10}\, \mathrm{M_\odot}$. Its H$\alpha$ kinematics exhibit overall ordered disk rotation, and its star formation rate (SFR) and effective radius ($R_{\rm eff}$) place it near the $M_\star-\rm SFR$ and $M_\star - R_{\rm eff}$ scaling relations at $z=2.2$ (\citep{genzel_rings_2008, Tacchella2015, Genzel2017, ForsterSchreiber2018}. Previous rotation curve analysis suggests it is strongly baryon-dominated and has a very massive compact central bulge mass component \citep{Genzel2017, Price2021_rc41, nestor_shachar_rc100_2023}. However, this galaxy has a very unusual structure: both \emph{HST} imaging and high-resolution H$\alpha$ emission maps show a clumpy ring-like structure with a clear lack of emission from the central region. It was believed that observations longer wavelengths would reveal a smooth disk with a central bulge, as was the case for similar galaxies in such redshifts. However, recent \emph{JWST}/NIRCam observations (including in the F444W filter at 4.4 $\rm {\mu m}$) show that the central region is dark even at rest-frame $1.4\mathrm{\mu m}$. Moreover, there is no detection of AGN activity \citep{Genzel2014}. This raises the possibility that the stellar content of this galaxy is indeed distributed in a clumpy ring-like structure.

Galactic rings have been studied extensively in the Local Universe, and are present in $\sim 20\%$ of spiral galaxies \citep[e.g.,][]{buta_galactic_1996}. Typically, these are photometric features associated with high star-formation, thus emitting mostly in rest-frame optical-UV, and they do not contain a significant fraction of the total mass in the galaxy \citep{Gerin1988, Garcia-Barreto1991, horellou_co_1995}. In disk galaxies, non-axisymmetric features such as bars and spiral arms lead to the formation of inner and outer rings, as torques drive gas to accumulate along Lindblad resonances \citep[e.g.][]{Goldreich1978_lindblad_resonance, Schwarz1981_rings_resonance, Byrd1994_resonance_rings_sim}. Such rings have been extensively observed in the Local Universe \citep{Buta1986_galactic_ring_obs, Buta1995_resonance_rings_obs, Byrd2006_resonance_rings_obs, Comeron2014_resonance_rings_cat, Abraham2025_rings}. However, such structures are typically faint and require a dominant non-axisymmetric component, missing in zC406690. Alternatively, under certain conditions, rings can form through mergers if one galaxy crosses through another galaxy. Such a ``pierce-through" creates a shockwave first moving radially inside, before bouncing back outwards as the gas settles in a ring \cite[e.g., The ``Cartwheel" galaxy, and][]{Lynds1976, Theys1977_rings_pierce, Toomre1978_piercethrough, Barnes1992_rings_overview, Higdon1995_cartwheel}.

At higher redshifts, detections of ring structures are less common, in part due to their low surface brightness and the limited resolution available, and in part because any axisymmetrical structures (i.e., bars) probably did not have much time to dynamically interact with the gas, as they are likely to have been formed recently (few $\sim \rm Gyr$) \citep{Espejo2025_morphologies}. Recent analytical models suggest the conditions at $z \sim 1-2$, with high gas fractions and high turbulence, are such that long-lived ring structures can form in massive ($\gtrsim 3 \times 10^{9}\, M_{\odot}$) SFGs, through accretion of aligned streams from the circum-galactic medium (CGM) \citep{dekel_origin_2020, Stern2024_CGMinflow}. In a marginally-stable disk the ring will quickly form SF clumps that are expected to lose angular momentum through torques \citep{Dekel2009}, therefore constant accretion is needed to sustain the ring. If rings are indeed massive and long-lived, they would leave their imprint on the gravitational potentials and the rotation patterns of the entire galaxy.

\begin{figure*}[t]
    \centering
    \includegraphics[width=0.95\linewidth]{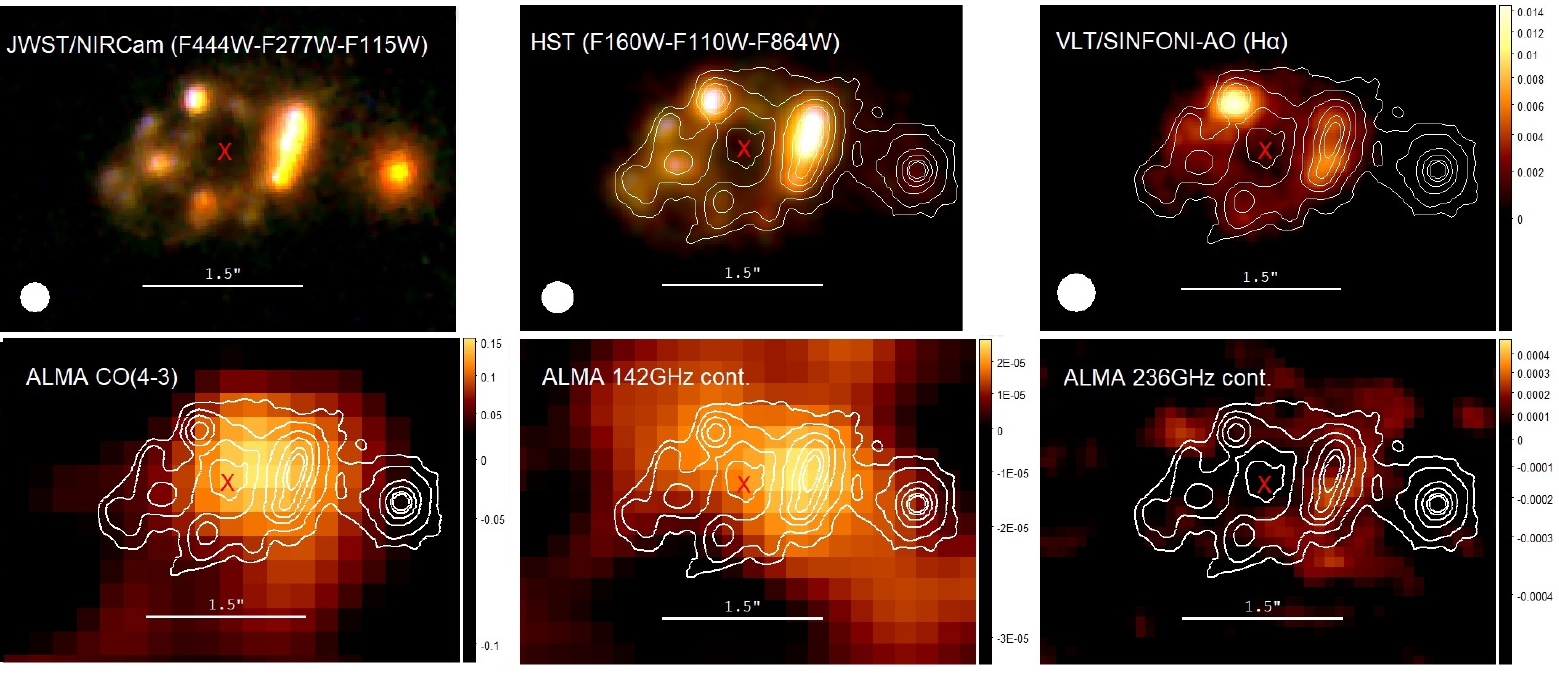}
    \caption{Imaging of zC406690 from various observations, showing a ring-like structure in spatially resolved images. The contours in all panels trace the \emph{JWST}/NIRCam F444W emission. The H$\alpha$ kinematic center is marked by the red cross. Top row: composite RGB image from NIRCam F115W, F277W and F444W (PSF matched to $0.15^{\prime\prime}$ using WebbPSF and Photoutils, left panel); composite RGB image from HST H, J and F814W bands (PSF $0.15^{\prime\prime}$, middle panel); Top right: H$\alpha$ moment zero map from VLT/SINFONI (PSF $0.18^{\prime\prime}$, right panel, in $10^{-17}$ Wm$^{-2}$). Bottom row: low-resolution ALMA archival observations of the CO(4-3) emission (PSF $1.5^{\prime\prime}$, integrated S/N$=7$, in Jy/beam km/s), 142GHz continuum (PSF $1.5^{\prime\prime}$, peak S/N$=4.5$, in Jy/beam) and 236 GHz continuum (PSF $0.74^{\prime \prime}$, peak S/N$=2.5$, in Jy/beam).}
    \label{fig:zC406690_flux_maps}
\end{figure*}
\begin{figure*}[t]
    \centering
    \includegraphics[width=0.75\linewidth]{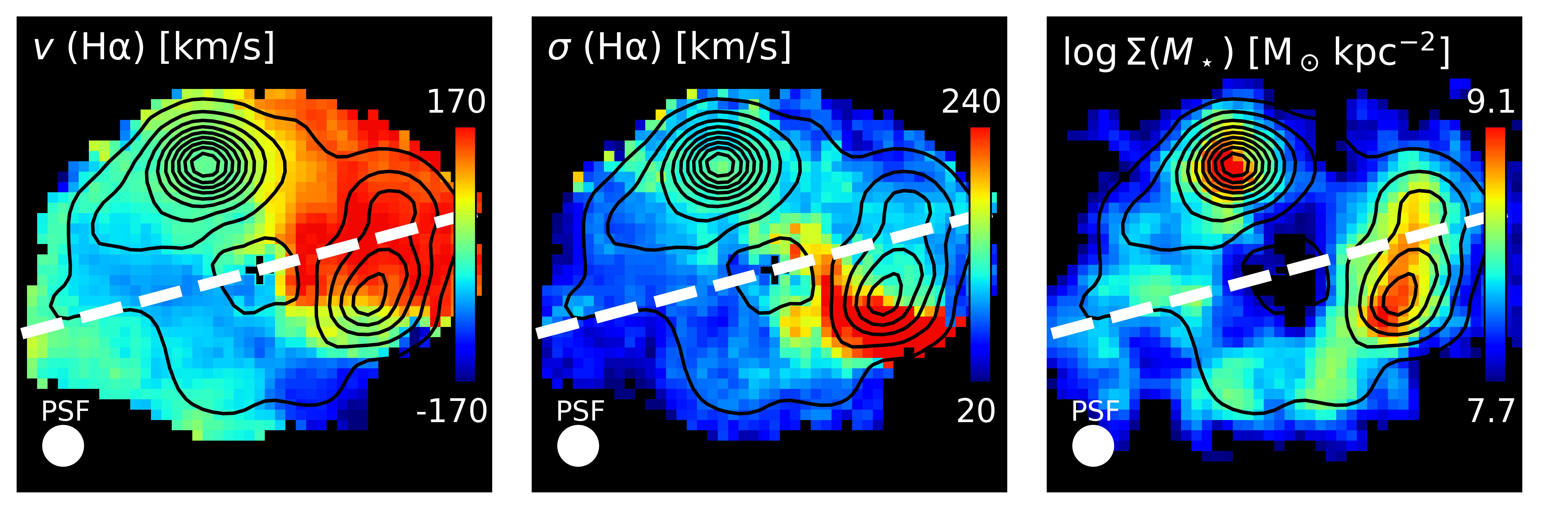}
    \caption{VLT/SINFONI adaptive-optics H$\alpha$ kinematics (PSF FWHM 0.18$^{\prime\prime}$) and \emph{HST} SED-fitting stellar maps from \citealt{ForsterSchreiber2018} (Figure 16). Contours trace the H$\alpha$ flux, and the the kinematic major axis ($PA=-70^o$) shown by the dashed white line. The strong twist in velocity and high $\sigma$ regions just south of the western clump is caused by the influence of strong outflows affecting the pixel-by-pixel single-Gaussian fit used to derive the kinematic maps. The blue-shifted broad emission broadens and shifts the line profile, increasing both centroid velocity and velocity dispersion estimates.}
    \label{fig:zc406690_sinfoni_nmfs2018}
\end{figure*}
In this paper, we present a new analysis of the unique ring galaxy zC406690 incorporating new constraints from recent high-resolution JWST/NIRCam imaging and archival low-resolution ALMA interferometry. We clearly detect the ring structure in all NIRCam filters, and identify a detection of the $\mathrm{CO}(4-3)$ line at $\sim$ 142GHz and dust continuum emission. We use these observations to estimate the molecular gas and dust masses of the galaxy, and compare the ionized and molecular gas kinematics. Our analysis supports the existence of a massive stellar and gas ring in this galaxy, with a highly extincted bulge in the center ($A_{\rm V} \sim 30$). In \S~\ref{sec:2.zc406690} we summarize the observed properties of zC406690; in \S~\ref{sec:3.methods} we define a razor-thin mass distribution in the shape of a ``Gaussian Ring" and discuss its properties; in \S~\ref{sec:4.fitting_procedure} we describe the mass modelling and fitting procedures for the rotation curve. We summarize our results in \S~\ref{sec:5.results}, and discuss possible scenarios in \S~\ref{sec:6.discussion}.

Throughout this paper, we assume a flat $\mathrm{\Lambda CDM}$ cosmology with $\Omega_{\mathrm{m}}=0.3$, $\Omega_{\mathrm{\Lambda}}=0.7$ and $H_0 = 70\, \mathrm{km\, s^{-1}\, Mpc^{-1}}$.

\section{Target: zC406690}\label{sec:2.zc406690}
zC406690 is a rotating, massive SFG at $z=2.196$ (RA 09:58:59.1, dec 02:05:04.2) with a plethora of observations  (See Figure~\ref{fig:zC406690_flux_maps}). It is included in the COSMOS and zCOSMOS surveys \citep{lilly_zcosmos_2007, scoville_cosmic_2007, Mancini2011} and in the SINS/zC-SINF surveys with the VLT/SINFONI \citep[with and without adaptive-optics, ][]{ForsterSchreiber2009, ForsterSchreiber2018}. Recently, it was observed with \emph{JWST}/NIRCam \citep{Casey2023}. Previously unpublished archival ALMA observations are also available (2013.1.00668.S, 2017.1.01020.S). We use the existing high-resolution H$\alpha$ spectroscopy from \cite{ForsterSchreiber2018} taken with SINFONI-AO to reanalyze the rotation curve, and provide new analysis for the properties of the cold gas based on band 4 ALMA data, for both the CO(4-3) emission line and dust continuum. The \emph{JWST}/NIRCam data is used to provide insight to the morphology of the galaxy and to constrain the dust attenuation models through the non-detection of a central emission in the reddest filters.

zC406690 is a near main sequence galaxy, with stellar mass $M_\star = 4 \times 10^{10}\, \mathrm{M_\odot}$, star-formation rate $\mathrm{SFR} = 300\, \mathrm{M_\odot \, yr^{-1}}$, and roughly half-solar metallicity, $12 + \log{\mathrm{O/H}} = 8.38$ based on the $\rm {[NII] / H\alpha} = 0.12$ line ratio. \citep{Mancini2011, ForsterSchreiber2018}. Its SFR is slightly elevated compared to the MS, by a factor of x3.5, but is still considered compatible. The ring contains several large star-forming clumps (clearly seen also in H$\alpha$), which have been studied extensively in \cite{genzel_sins_2011} and \cite{newman_shocked_2012}. The brightest clump structures have individual SFRs of 10-40 $\mathrm{M_\odot\, yr^{-1}}$ (or $\Sigma_{\mathrm{SFR}}=2-13\, \mathrm{M_\odot\, yr^{-1}\, kpc^{-2}}$) and, based on $\rm SFR - M_\star$ scaling relations, collectively contain a molecular gas mass of $M_{\mathrm{mol,clumps}} \sim 3.4 \times 10^{10}\, \mathrm{M_\odot}$. These clumps drive ionized gas outflows of several hundred $\mathrm{km\, s^{-1}}$ which could rapidly deplete their gas content within $\lesssim 100\, \mathrm{Myr}$ \citep{genzel_sins_2011, newman_shocked_2012}. We summarize the basic properties of zC406690 in Table~\ref{tab:zc406690_properties}.

\begin{table}[]
\resizebox{\columnwidth}{!}{%
\centering
\begin{tabular}{c|c|c}
property & value & reference \\ \hline \hline
$z$ & $2.196$ & [1, 2]\\
$M_\star \left[ 10^{10} \mathrm{M_\odot} \right]$& $4.14$ & [1, 2]\\
$i \left(^{o} \right)$ & $26^{+15}_{-5}$&  [3]\\
$r_{\mathrm{1/2, H\alpha}} \left[ \mathrm{kpc} \right]$& $4.8 \pm 0.1$& [2]\\
$F_{\mathrm{H\alpha}} \left[ 10^{-16}\, \mathrm{erg\, s^{-1}\, cm^{-2}} \right]$& $3.01 \pm 0.05$&[2]\\
$SFR \left[ \mathrm{M_\odot\, yr^{-1}} \right]$& $300$& [1, 2]\\
 $\delta SFR_{\mathrm{MS}}\ ^a$& $0.55$&-\\
$\mathrm{ [NII] / H\alpha}$ & $0.12 \pm 0.01$& [2]\\
$12 + \log {\left( \mathrm{O/H} \right) }$ & $8.38 \pm 0.08$& [2]\\
$M_{\mathrm{mol, clumps}} \left[ 10^{10} M_\odot \right]$& $3.4^{+1.2}_{-1.5}$ & [4, 5]\\
 $V_{\mathrm{circ}} \left[ \mathrm{km\, s^{-1}} \right]$& $276 \pm 55$&[3, 6, 7]\\
 $\sigma_0 \left[ \mathrm{km\, s^{-1}} \right]$& $70 \pm 5$ &[3, 6, 7]\\
 $V_{\mathrm{circ}} / \sigma_0$& $3.8 \pm 0.8$&-\\
\multicolumn{1}{c|}{$M_{\mathrm{baryon}} \left[ 10^{10} \mathrm{M_\odot} \right]$} & \multicolumn{1}{c|}{$11.07 \pm 0.05$} & [3, 6, 7]\\
bulge-to-total (B/T) & $0.9 \pm 0.07$ & [3, 6, 7]\\ 
\end{tabular}%
}
\caption{Selected properties of zC406690, and their references. $^a$Deviation from the SFR-$M_\star$ main sequence, $\delta SFR_{\mathrm{MS}} = \log {\left( SFR / SFR_{\mathrm{MS}} \right)}$, from \cite{Whitaker2014a}. $^1$\cite{Mancini2011}, $^2$\cite{ForsterSchreiber2018} $^3$\cite{Genzel2020}, $^4$\cite{genzel_sins_2011}, $^5$\cite{newman_shocked_2012}, $^6$\cite{Price2021_rc41}, $^7$\cite{nestor_shachar_rc100_2023}.}
\label{tab:zc406690_properties}
\end{table}

\begin{figure*}[t]
    \centering
    \includegraphics[width=0.8\linewidth]{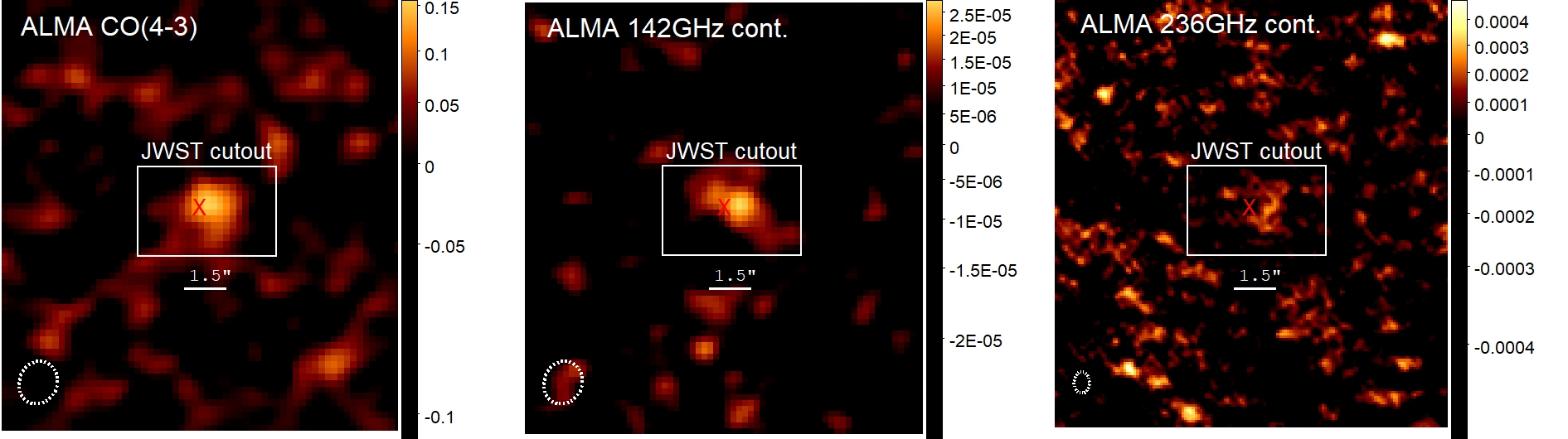}
    \caption{A larger field of view (FOV) for the CO(4-3) line (left, $\rm {S/N}=7$), 142GHz continuum (middle, $\rm {S/N}=4.5$) and 236GHz continuum (right, $\rm {S/N}=2.5$) obtained with ALMA The PSF size is shown by the ellipse in the bottom left corner. The white rectangle marks the NIRCam cutout shown in Figure~(\ref{fig:zC406690_flux_maps}) and the red cross marks the H$\alpha$ kinematic center. All colorbars are units of $\rm Jy/beam$.}
    \label{fig:zC406690_flux_maps_zoomout}
\end{figure*}
The H$\alpha$ spectroscopy from \cite{ForsterSchreiber2018}, shown in Figure~(\ref{fig:zc406690_sinfoni_nmfs2018}), allow for the construction of an extended rotation curve along the kinematic major-axis, up to twice the effective radius. We base our kinematic analysis in sections \ref{sec:4.1mass modelling} and \ref{sec:4.2RC fitting} on the reduced data and kinematic PA derived in \cite{ForsterSchreiber2018}, which are described in detail in their paper. In short, 10 hours of data was collected for this source during 2010-2011 with the VLT/SINFONI including adaptive-optics, at a scale of $\rm 50\, mas/pixel$ and an effective spatial resolution of 0.18$^{\prime \prime}$ for the reduced data cubes. The kinematic maps were obtained by a pixel-by-pixel single-Gaussian fitting (with H$\alpha$ $\rm S/N>5$ per pixel), with the kinematic center defined by the position with the largest velocity gradient and the kinematic PA as the direction of the line passing through the kinematic center maximizing the velocity difference.

These observations reveal an ordered rotation pattern with a strong drop in rotation velocity past $R_{\mathrm{eff}}$. Previous Kinematic fitting of yielded an extremely baryon-dominated galaxy within $R_{\mathrm{eff}}$, with DM fractions $f_{\mathrm{DM}}(R_{\mathrm{eff}}) = 0.06$ \citep{Genzel2020, Price2021_rc41, nestor_shachar_rc100_2023}. In particular, as mentioned, these models require most ($\sim$90\%) of the total baryon mass to be concentrated in a $9 \times 10^{10}\, \mathrm{M_{\odot}}$ central component. No such component is seen in stellar light probed with current optical to near-infrared imaging, in particular down to a surface brightness of 23.7 $\rm {mag/arcsec}^2$ ($5\sigma$) at 4.4$\rm {\mu m}$ (\emph{JWST} F444W band). However, these models follow a standard mass decomposition procedure, assuming the majority of the mass is distributed smoothly. In this work, we also explicitly take into account the observed ring-like morphology.

The dust and cold gas of zC406690 were previously observed with ALMA in band 4 and band 6 (programs 2013.1.00668.S, 2017.1.01020.S), but remained unpublished so far. We analyze the data and reveal a marginally-resolved detection of the CO(4-3) line emission (integrated $\rm{S/N} = 7$), as well as dust continuum in both 142GHz (rest 453GHz, $\rm{S/N} = 4.5$) and 236GHz (rest 754GHz, $\rm{S/N} = 2.5$). We reduced the data with the standard \texttt{scriptForPI} scripts in the corresponding CASA versions 4.2.2 (for program 2013.1.00668.S) and 5.1.1 (for program 2017.1.01020.S). Then, we run the \texttt{mstransform} task in CASA 5.7.2 to convert the two datasets onto the same spectral axis with a channel width of 25 km/s. Next, we recompute the visibility weights with \texttt{statwt} and subtract the continuum with \texttt{uvcontsub} using line-free channels. Finally, we run the \texttt{tclean} task to produce the clean data cube at the phase center 'J2000 149.746250$^\circ$ 2.084444$^\circ$. The clean threshold is set to 2-sigma, where sigma is the RMS noise measured from a previous clean run. Since the target is not highly spatially resolved, we do not use a complex clean mask or multiscale clean. The final CO(4-3) data cube for is then used for our analysis.

The flux maps for the reduced ALMA cubes are shown in the bottom row of Figure~(\ref{fig:zC406690_flux_maps}), and again in Figure~(\ref{fig:zC406690_flux_maps_zoomout}) with a wider FOV to compare with the background noise-level. The spatial resolution of these observations is too low to localize the origin of the emission (PSF FWHM $1.5^{\prime \prime}$ in band 4, and $0.74^{\prime \prime}$ in band 6). The centroid of the CO(4-3) emission, as well as the 142GHz continuum, peak at a location between the dynamical center and the western clump structures ($\approx 0.5^{\prime \prime}$ west of the center), yet the continuum emission seem to extend north-east to south-west, and not towards the dynamical center, possibly indicating that the dust follows the clumps. A specific determination as to the location of the dust is unclear given the low resolution, but the total gas and dust content of the galaxy is discussed in section~\ref{sec:5.3.Bulge_exntinction}. Furthermore, we extract a velocity map from the CO(4-3) emission using single-gaussian fitting for each spaxel, along with a few uncorrelated data points along the kinematic major-axis, which we compare with the H$\alpha$ kinematics in section~\ref{sec:5.2CO_kinematics}.


\subsection{Western Companion}
\begin{figure}[h]
    \centering
    \includegraphics[width=\columnwidth]{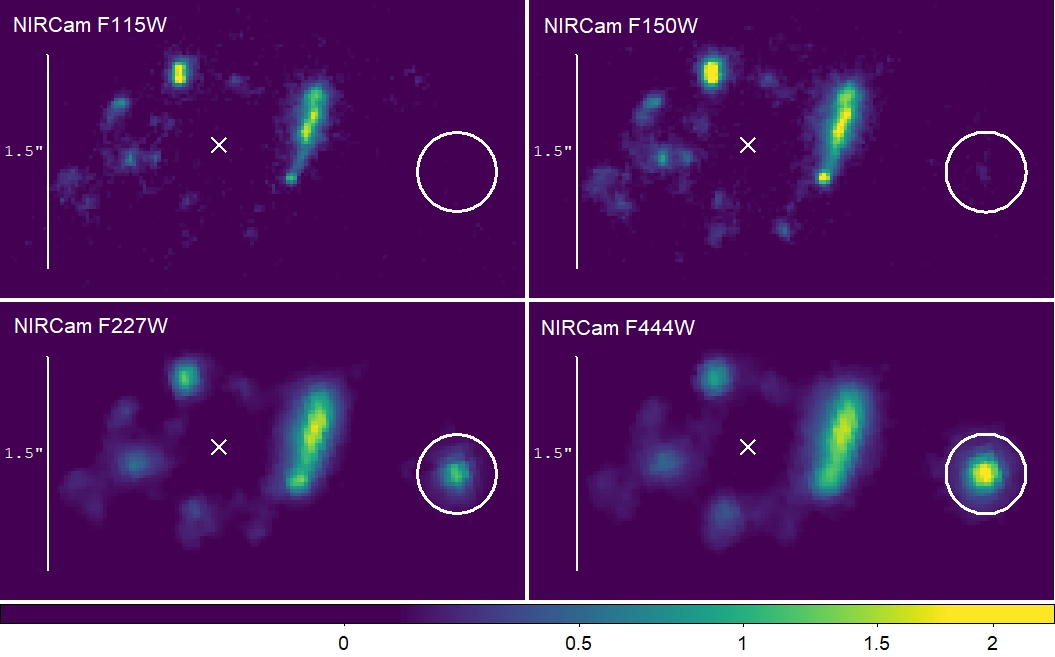}
    \caption{\emph{JWST} imaging showing the object west of zC406690, in F115W, F150W, F277W, F444W (from top-left to bottom-right), given in $\rm MJy/sr$. The circle shows the location of the companion candidate based on the F444W band, and the cross marks the kinematic center.}
    \label{fig:companion_jwst}
\end{figure}

A nearby compact red source is present to the west of the galaxy, about $1.6^{\prime \prime}$ west from its center and $1^{\prime \prime}$ west of the western clumps structure. It is virtually undetected in the bluest HST bands but becomes increasingly bright at wavelengths $>1\, \mathrm{\mu m}$ (see Figure~\ref{fig:companion_jwst}). 
No emission lines are detected in the available spectroscopic data, so no accurate spectroscopic redshift measurement can be measured. Based on Spectral Energy Distribution (SED) analysis using all avaialabe filters, we estimate its photometric redshift $z_\mathrm{phot} = 2.14^{+0.32}_{-0.42}$ and total stellar mass $\log M_\star / M_\odot = 10.25^{+0.16}_{-0.17}$, with a V-band extinction of $A_{\mathrm{V}} = 1.50^{+0.28}_{-0.24}$ (see Appendix \ref{app:SED_fitting}). Given the wide uncertainty in redshift, it remains unclear if it is indeed related to zC406690. 

One could speculate that if this is indeed physically associated, it could have an effect on the kinematics of zC406690. The sphere of influence for tidal forces can be estimated by the Jacobi radius \citep[see also][]{Genzel2017}, 
\begin{equation*}
R_J / R_{12} = \frac{1}{4} \left( \frac{M_2 / M_1}{0.05} \right)^{1/3}    
\end{equation*}
where $R_{12}$ is the separation between the smaller system $M_2$ and the larger system $M_1$ \citep[see][]{Genzel2017}. Given the stellar mass ratios of 2.25:1, this yields $R_J / R_{12} = 0.5$ implying that it should leave a trace on the western side of the velocity field of zC406690. The apparent kinematic perturbations seen in Figure~(\ref{fig:zc406690_sinfoni_nmfs2018}) to the south of the western clump complex are likely not due to an interaction but are associated with strong outflow signatures in H$\alpha$ and $\rm {[NII]}$ emission, skewing the kinematic map when a single-component fit is used at each pixel \citep{newman_shocked_2012, ForsterSchreiber2018}. The kinematics along the major axis (indicated by the solid line in Figure~\ref{fig:zc406690_sinfoni_nmfs2018}) show the smooth and symmetric variations expected for regular disk rotation \citep{Genzel2014, Genzel2017, Price2021_rc41, nestor_shachar_rc100_2023}.

Another possibility is that the companion pierced the center of zC406690 and is moving west, making zC406690 itself an interaction-induced ring galaxy with radial motions \citep[e.g.,][]{Lynds1976}. Without spectroscopy we cannot measure its peculiar velocity, however we can crudely estimate a time frame for such an interaction. Assuming a velocity of $\sim 200 \mathrm{km\, s^{-1}}$ and a relative angle of $45^o$ for a $1.5^{\prime\prime}$ separation, this violent interaction may have occurred $100-500\, \mathrm{Myr}$ ago, a few rotational dynamical timescales. In such scenario, one would expect to see a trail of stripped gas yet there is no such detection in any of the available spectroscopic data. Without spectroscopic confirmation of the physical association between zC406690 and this neighbour, we deem any discussion to be too speculative and do not consider this scenario further.

\section{Ring Mass Distribution}\label{sec:3.methods}
A massive stellar ring creates its own gravitational potential, very different than that of an exponential disk, and would have to be explicitly considered when calculating the circular velocity of the galaxy. In the following section we propose an analytical model for a razor-thin axisymmetric ring, with a radial Gaussian surface density profile which is offset from the center. Even though an intrinsic thickness is to be expected in a galaxy at $z \sim 2$ (as we discuss in \S~\ref{sec:1.intro}), an infinitely thin ring is an adequate approximation for a toroidal mass distribution, particularly along the mid-plane, with deviations of less than $\lesssim 1 \%$ in the calculation of the potential \citep{Hure2019, Hure2020}. We therefore discuss only the 2D and determine its main properties in the following section.

\subsection{Razor-Thin Gaussian Ring profile} \label{sec:3.1Gaussain Ring}
For a ring with total mass $M_{\mathrm{ring}}$, the Gaussian ring is defined by three main parameters: a peak radius $R_{\mathrm{p}}$, standard deviation $\sigma_{r}$ and a characteristic surface density $\Sigma_0$. The surface density is a Gaussian shifted by $R_{\mathrm{p}}$:
\begin{equation} \label{eq:gaussian_ring_surface_density}
    \Sigma(r) = \Sigma_0 \times \exp\left\{-\frac{(r-R_{\mathrm{p}})^2}{2\sigma_{r}^2}\right\}
\end{equation}
\begin{figure}[h]
    \centering
    \includegraphics[width=\columnwidth]{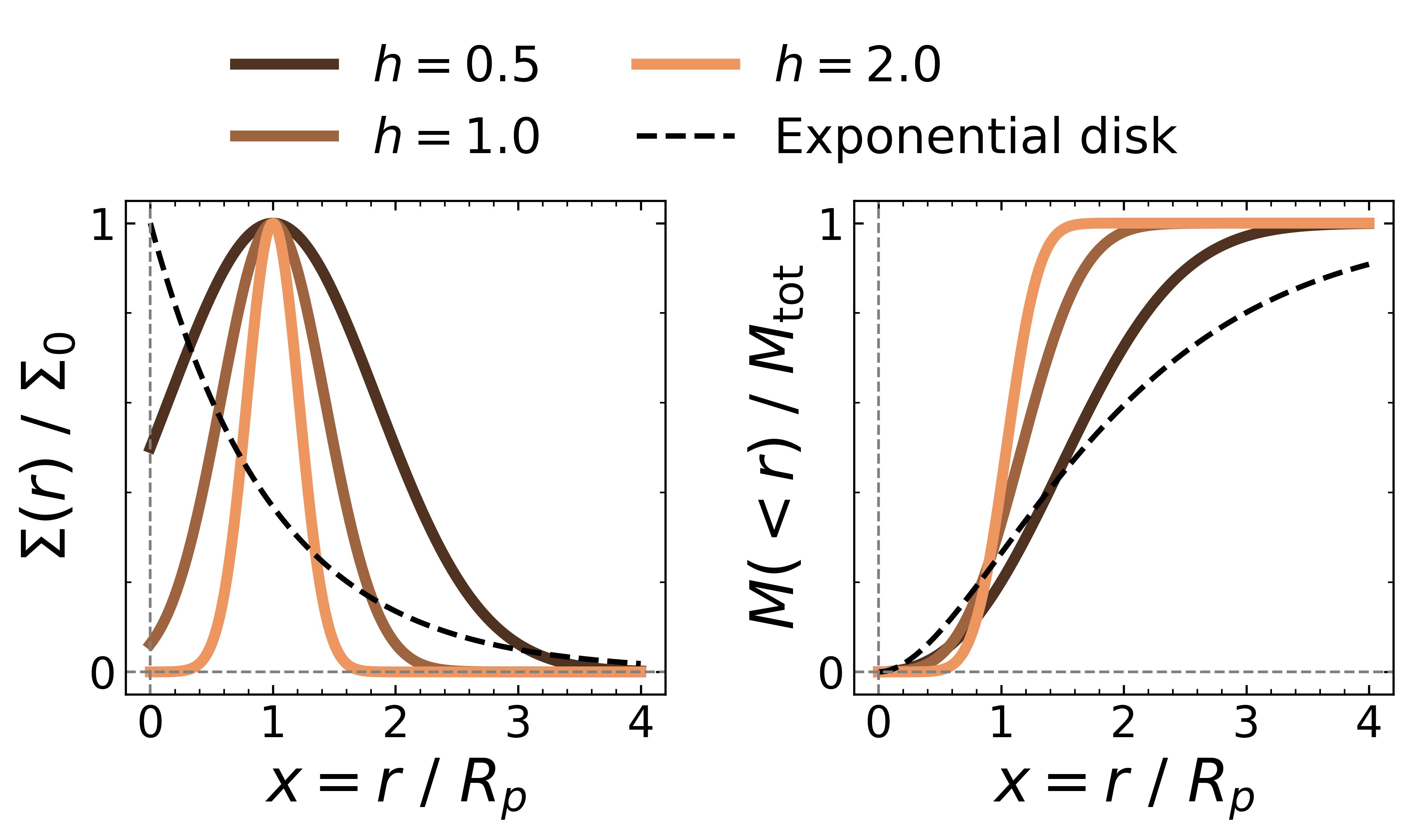}
    \caption{Surface density $\Sigma(r)$ (left) and cumulative mass $M(<r)$ (right) as a function of the normalized radius $r/R_{\mathrm{p}}$ for different shape parameters of a Gaussian Ring. The dashed line shows a razor-thin Exponential disk for comparison \citep{Freeman1970}.}
    \label{fig:GaussianRing_dimless_densityMass}
\end{figure}
The full-width at half-maximum is $F_r = \sqrt{8\ln{2}}\, \sigma_r$. We define the ``shape parameter" as the ratio of the peak radius to $F_r$, to distinguish ``narrow" versus ``wide" rings:
\begin{equation}\label{eq:shape parameter}
    h = \frac{R_{\mathrm{p}}}{F_r} \approx 0.42 \frac{R_{\mathrm{p}}}{\sigma_{\mathrm{r}}} \ \ \ \ 
    \begin{cases}
        \mathrm{narrow} & h \gg 1 \\
        \mathrm{wide} & h \ll 1
    \end{cases}
\end{equation}
We expect the shape parameter to vary between $0.1 < h < 10$, and typically be $\sim O(1)$ \citep[motivated by][]{dekel_origin_2020}, as very narrow rings are less likely to be stable. We take $h=1$ as our fiducial model.  

\begin{figure}[h]
    \centering
    \includegraphics[width=\columnwidth]{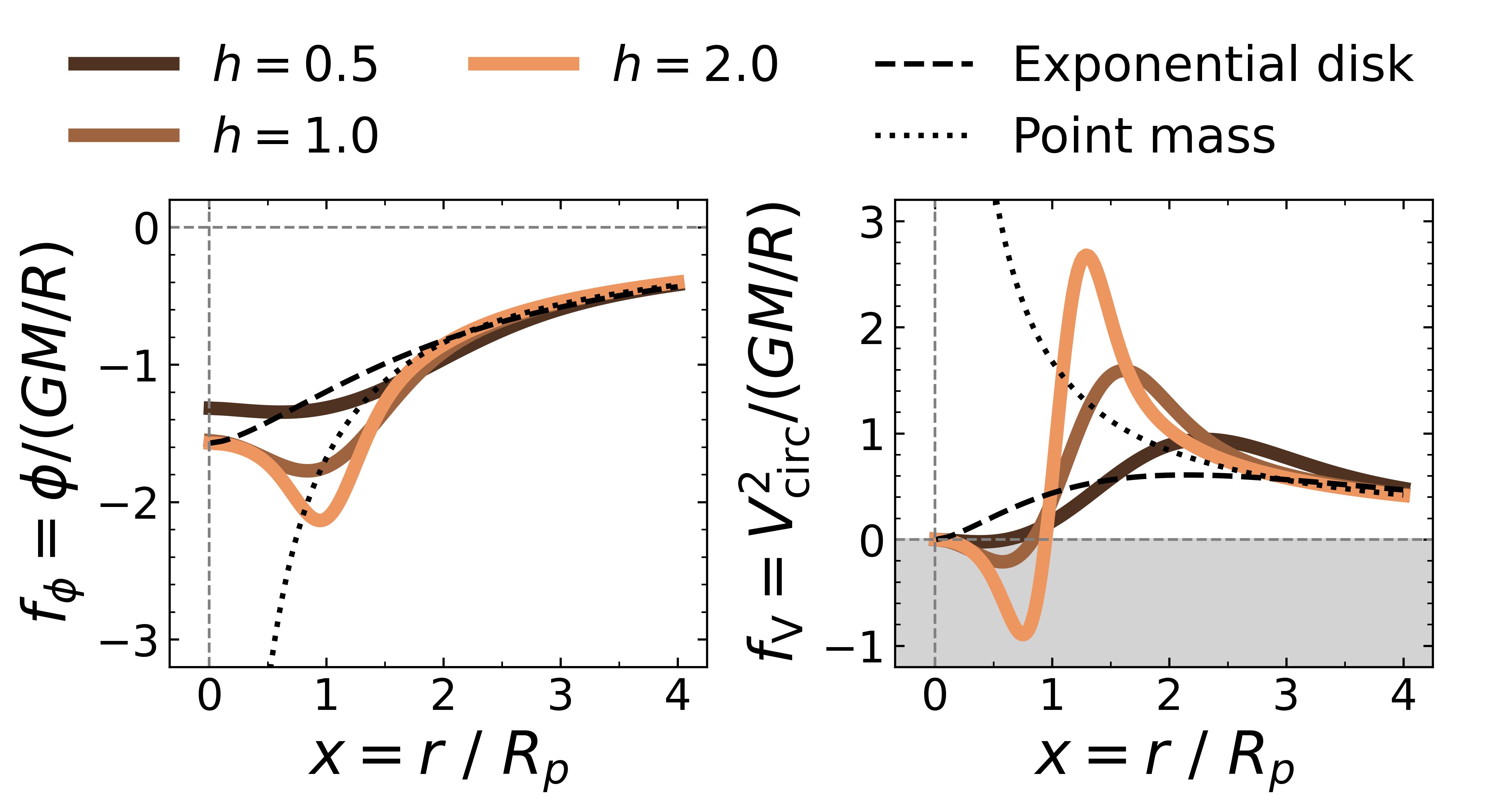}
    \caption{Normalized potential $\phi$ (left) and Normalized circular velocity $V^2_{\mathrm{circ}}$ (right) as a function of the normalized radius $r/R_{\mathrm{p}}$ for different shape parameters of a Gaussian Ring. The dashed line shows a razor-thin exponential disk for comparison, and the dotted line a point mass \citep{Binney1987}. The shaded grey area is unstable for rotational support under the ring's self-gravity.}   \label{fig:GaussianRing_dimless_potentialVelocity}
\end{figure}

The gravitational potential and circular velocity can be derived analytically \citep[following][their Eq. (2.155)]{Binney1987}. We refer the reader to Appendix~\ref{app:GaussianRing} for the full derivation of the expressions. We derive dimensionless functions depending solely on $h$, such that the total mass $M_{\rm ring}$, gravitational potential $\phi(r)$ (in the plane of the ring), and the circular velocity $V^2_{\mathrm{ring}}(r)$ are:
\begin{equation}\label{eq:gaussian_ring_potentials}
\begin{split}
    M_{\mathrm{ring}} &= 2 \pi R_{\mathrm{p}}^2 \Sigma_0 \times f_{\mathrm{M}} (h) \\
    \phi(r) & = \frac{G M_{\mathrm{ring}}}{R_{\mathrm{p}}} \times f_{\phi}(\mathrm{r}) \\
    V^2_{\mathrm{ring}}(r) & = \frac{G M_{\mathrm{ring}}}{R_{\mathrm{p}}} \times f_{\mathrm{V}}(\mathrm{r}) \ \ \ 
\end{split}
\end{equation}
with $f_V = -r \frac{d f_{\phi}}{dr}$. The dimensionless functions $f_{\mathrm{V}}$, $f_{\phi}$ and $f_{\mathrm{M}}$ fully determine the shape of the profile, and are a function of the shape parameter $h$ alone (see Figure~\ref{fig:GaussianRing_dimless_potentialVelocity}, and appendix~\ref{app:GaussianRing_general_equations} for more details). The ring's mass and peak radius (or FWHM, given a value of $h$) only affect the amplitude. For reference, a point mass will have $f_{\phi} = - R_{\mathrm{p}}/r$ and $f_{\mathrm{V}} = R_{\mathrm{p}}/r$. 

The effective radius, $R_{\mathrm{eff}}$, of the Gaussian ring is always larger than the peak radius. It increases for wider rings (small $h$) and approaches the peak radius as the ring becomes narrower (large $h$), as a function of the shape parameter alone. We find that the ratio of the effective radius to the peak radius is well approximated for $0.1 \leq h \leq 10$ (to within 6\% accuracy) by the formula (see Appendix \ref{app:GaussianRing_general_equations}):
\begin{equation}\label{eq:GaussianRing_Reff}
    R_{\mathrm{eff}} / R_{\mathrm{p}} \approx 1 + \frac{1}{4} h^{-5/4} \ \ \ .
\end{equation}
For our fiducial $h = 1$, $R_{\mathrm{eff}} \approx \frac{5}{4} R_{\mathrm{p}}$. In the limiting case of an infinitely narrow ring (delta function), $h \rightarrow \infty$, we get $R_{\mathrm{eff}} = R_{\mathrm{p}}$, as expected. \\ 
The circular velocity is well-defined as long as the dimensionless function $f_{\mathrm{V}}$ is positive at all radii. However, due to the central cavity of this mass distribution, there must be a region within some radius $r_{\mathrm{circ}}$  ($r_{\mathrm{circ}} < R_{\mathrm{p}}$) where the net force acting on a test particle will always be pointing outwards, and centrifugal equilibrium is impossible. We define $r_{\mathrm{circ}}$ as the radius where the potential gradient becomes positive and $V^2_{\mathrm{circ}}$ becomes non-negative, i.e., by solving for $f_{\mathrm{V}} (r_{\mathrm{circ}}) = 0$ in Equation~(\ref{eq:gaussian_ring_potentials}). The value of $r_{\mathrm{circ}}$ is a function of the shape parameter $h$ alone (see Appendix \ref{app:GaussianRing_circularEquilibrium}), and for our fiducial $h=1$ we find $r_{\mathrm{circ}} \approx 0.7\, R_{\mathrm{peak}}$. Thus, a Gaussian ring is never a rotationally stable system on its own, and requires an additional mass component to support its self-gravity. \\

To maintain centrifugal equilibrium across all radii, we consider an additional central mass component that sufficiently deepens the potential to ensure a net inward-force. This is also motivated by the fact that ring structures are indeed often found together with central stellar bulges \citep{buta_galactic_1996, genzel_rings_2008}. We thus introduce a second component with total mass $M_{\mathrm{bulge}}$, in the form of either a central point mass or a Sérsic bulge (with $n_{s} = 4$ and $R_{\mathrm{eff,bulge}}$), and look for the minimal $M_{\mathrm{bulge}}$ that enables rotational equilibrium at all radii (i.e., satisfying Eq.~\ref{eq:Gaussian_ring_stabilizing_criteria}). We find that for our fiducial model $h=1$, the point mass must be at least $ > 0.1 M_{\mathrm{ring}}$ whereas the mass of a Sérsic bulge must be $ \gtrsim 0.1-0.3 M_{\mathrm{ring}}$, depending on its size (see Appendix \ref{app:GaussianRing_fluxRing}). Generally, narrower rings require a larger stabilizing mass (up to $\lesssim M_{\mathrm{ring}}$), depending on the assumed bulge radius.  This acts as a lower limit in our kinematic modeling, to ensure kinematic stability at all radii.

\subsection{Dynamical vs.~Flux ring}\label{sec:3.3light_weighting}
Due to highly localized SFR activity, ring-like morphologies may appear even when the bulk of the underlying mass is more smoothly distributed, such as in an extended disk configuration. This has been previously reported from \emph{HST} imaging when incorporating near-IR imaging bands \citep{Wuyts2012, Lang2014, Tacchella2015} and has become even clearer with current longer wavelength \emph{JWST} imaging \citep[e.g.][]{ferreira_jwst_2023, jacobs_early_2023}. Common spectroscopic tracers (such as $\mathrm{H\alpha}, \mathrm{\left[OIII \right]}, \mathrm{\left[NII \right]}$, etc.) are mostly emitted from $\mathrm{HII}$ regions typically surrounding massive OB stars near regions with high SFRs, and will therefore be biased against older stellar populations which are more abundant in central bulges, for example. Thus the optical-NIR wavelengths trace more closely the stellar mass distribution being less affected by star formation and dust effects.

\begin{figure}[h]
    \centering
    \includegraphics[width=\linewidth]{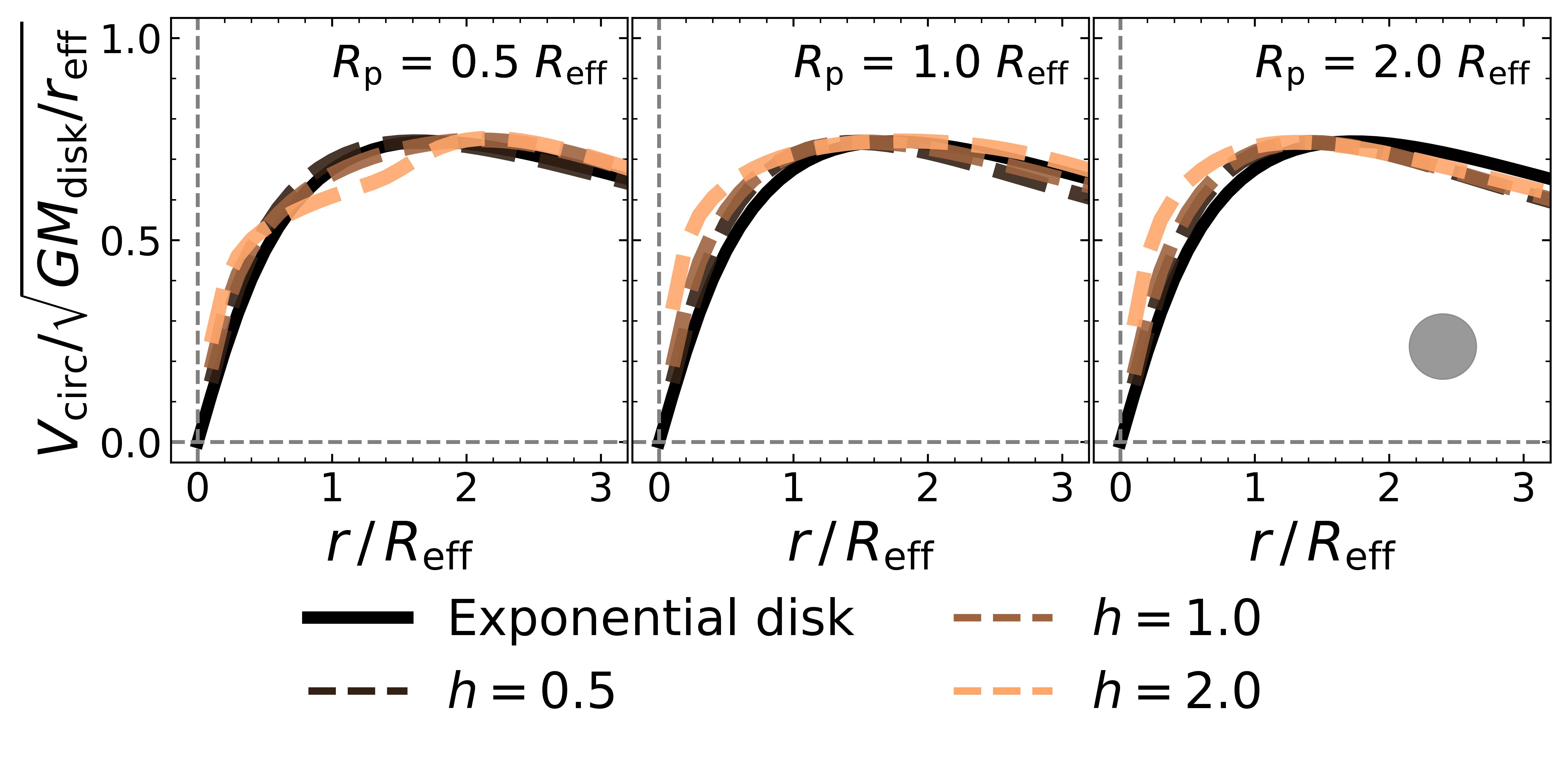}
    \caption{The effect of a ``flux ring" on the observed circular velocity, for a pure exponential disk. The black curve shows the circular velocity for an exponential disk. The dashed colored lines show the observed rotation curve for the same disk, except that the ring is dark and light traces a ring given by Eq.~(\ref{eq:gaussian_ring_surface_density}). Colors match different ring shape parameters, $h = 0.5, 1, 2$. Each panel represents a different location of the ring's peak with respect to the effective radius of the disk: inside (left), coinciding (middle) and outside (right). The disk mass is $M = 10^{10}\, M_{\mathrm{\odot}}$ with $R_{\mathrm{eff}} = 4\, \mathrm{kpc}$, the inclination is $25^{\mathrm{o}}$ and the PSF FWHM is $0.5 R_{\mathrm{eff}}$.}
    \label{fig:flux_ring_effects}
\end{figure}

An observed luminous ring structure may not trace the actual underlying mass distribution. But by examining the kinematics, it is possible to distinguish between a true mass enhancement (``dynamical ring") versus an emission/extinction effect in an otherwise typical smooth disk distribution (``flux ring"). In observations with a limited spatial resolution, where the $PSF \sim R_{\rm eff}$, the inferred rotation curves are greatly affected by regions with increased tracer emission, an effect commonly referred to as ``flux weighting". Figure \ref{fig:flux_ring_effects} shows how a ``flux ring" affects the observed RC of a pure exponential disk. Each panel shows the effect of changing the shape parameter of the ring (the brighter the narrower), for peak radii, $R_{\mathrm{p}}$, smaller, equal to, or larger than the disk effective radii $R_{\mathrm{eff}}$. The effect is most significant when $R_{\mathrm{p}} \gtrsim R_{\mathrm{eff}}$, impacting the shape of the RC and the location where the velocity flattens and starts to drop (``turnover radius"). Of course, as this is a projection effect, it will become more significant as the inclination is more face-on and as the beam size increases (see Appendix~\ref{app:GaussianRing_fluxRing}).

\section{Fitting procedure}\label{sec:4.fitting_procedure}
\subsection{Mass modelling} \label{sec:4.1mass modelling}
We use a parametric forward-modelling approach, with a mass model consisting of a dark matter halo, a central stellar bulge, a thick disk and a razor-thin Gaussian ring (see \S~\ref{sec:3.1Gaussain Ring}). The thick disk has an exponential profile (with a Sérsic index of $n_{\mathrm{s}}=1$) with a total mass $M_{\mathrm{disk}}$, effective radius $R_{\mathrm{eff,disk}}$ and an intrinsic axis-ratio $q$. The bulge is spherical with a de Vaucouleurs profile (Sérsic index $n_{\mathrm{s}}=4$), a total mass $M_{\mathrm{bulge}}$ and an effective radius of $R_{\mathrm{eff,bulge}} =1\, \mathrm{kpc}$ to represent a compact distribution. The dark matter halo is also spherical, with mass $M_{\mathrm{DM}}$ and concentration parameter $c_{\mathrm{NFW}}$ \citep{Navarro1996}. \\
The circular velocity is the sum in quadrature over all components:
\begin{equation}
    V_{\mathrm{circ}}^2 = V_{\mathrm{DM}}^2 + V_{\mathrm{bulge}}^2 + V_{\mathrm{disk}}^2 + V_{\mathrm{ring}}^2
\end{equation}
where $V_{\mathrm{i}}$ are the circular velocities of the dark matter halo, bulge, disk and ring components. 

The circular velocities of the bulge and disk are calculated following \cite{Noordermeer2008} for Sérsic profiles, taking into account the intrinsic thickness. The Sérsic profile is defined by the surface density:
\begin{equation}\label{eq:Sersic_profile}
    \Sigma(r)= \Sigma_0  \exp {\left\{ -\left(r/R_{\mathrm{d}} \right)^{1/n} \right\}}
\end{equation}
where $R_{\mathrm{d}}$ represents the scale radius of the disk, and the total mass is $M_{\rm{s \acute{e} rsic}} = 2\pi R_{\mathrm{d}}^2 \Sigma_0 n \Gamma(2n)$.

For the DM halo we assume a generalized NFW profile with a varying inner slope \citep[``$\alpha$NFW",][]{Navarro1996}:
\begin{equation}\label{eq:NFW_profile}
    \rho = \frac {\rho_0}{\left(r/r_s\right)^{\alpha} \left(1+r/r_s\right)^{3-\alpha}}
\end{equation}
with $\alpha = 1$ for the standard NFW profile, which is the default choice unless stated otherwise. The circular velocity is then simply the Keplerian velocity $V_{\mathrm{DM}} = \sqrt{GM(<r)/r}$. 

In addition, gas motions are affected by pressure gradients, that alter the equilibrium rotation velocities. Assuming hydrostatic equilibrium, the corrected rotational velocity is given by \citep{Burkert2010}:
\begin{equation}
\label{eq:burkert}
    V_{\mathrm{rot}}^2 = V_{\mathrm{circ}}^2 + \sigma_0^2 \frac{d \ln \rho}{d \ln r}
\end{equation}
taking a constant velocity dispersion $\sigma_0$ throughout the midplane, as it has been shown to be a good approximation for MS SFGs at $z\sim 1-2$ \citep{genzel_rings_2008, Wisnioski2015, ForsterSchreiber2018}. zC406690 exhibits a similar close-to-constant velocity dispersion, with the exception of strong outflows coming from the clump structures \citep[][and see Figure~\ref{fig:zc406690_sinfoni_nmfs2018}]{genzel_sins_2011}. Small radial variations in velocity dispersion have little effect on the resulting $v_{\rm rot}$, weakening the effect of the pressure support.

As long as the surface density decreases with radius, the rotation velocity will be reduced. For an exponential disk, Eq. (\ref{eq:burkert}) yields:
\begin{equation*}
    V_{\mathrm{rot}}^2 = V_{\mathrm{circ}}^2 - 2 \sigma_0^2 \left( \frac{r}{R_\mathrm{d}} \right)    \ \ \ 
\end{equation*}
and the rotational velocity is reduced at large radii. 
However, for a razor-thin Gaussian ring, the same correction term will have a dramatically different effect: within the ring the density gradient is positive thus increasing the rotation velocity, and vice versa. Inserting Eq.~(\ref{eq:gaussian_ring_surface_density}) into Eq.~(\ref{eq:burkert}) we find: 
\begin{equation}\label{eq:burkert_gaussian_ring}
    V_{\mathrm{rot}}^2 = V_{\mathrm{circ}}^2 - 8 \ln{2}\, h^2 \sigma_0^2 \times \frac{r(r-R_p)}{R_p^2} \ \ \ .
\end{equation}
Beyond the peak radius, the steep gradients will cause the circular velocity to drop sharply (as long as $\sigma_0 \sim \mathrm{const.}$), as the correction term scales with $r^2$. Thus, rapid declines in the rotation curves beyond the scale of the ring can also be driven by pressure gradients in a massive, dynamic ring. For example, for our fiducial model with $h=1$, even at $V_{\mathrm{circ}} / \sigma_0 = 5$ the velocity drops by as much as $\sim 45\%$ at $r = 2 R_{\mathrm{p}}$ (compared to $\sim 15\%$ for an exponential disk at $r = 2 R_{\mathrm{d}}$; or $30\%$ for a point mass).
\begin{figure}[h]
    \centering
    \includegraphics[width=0.7\columnwidth]{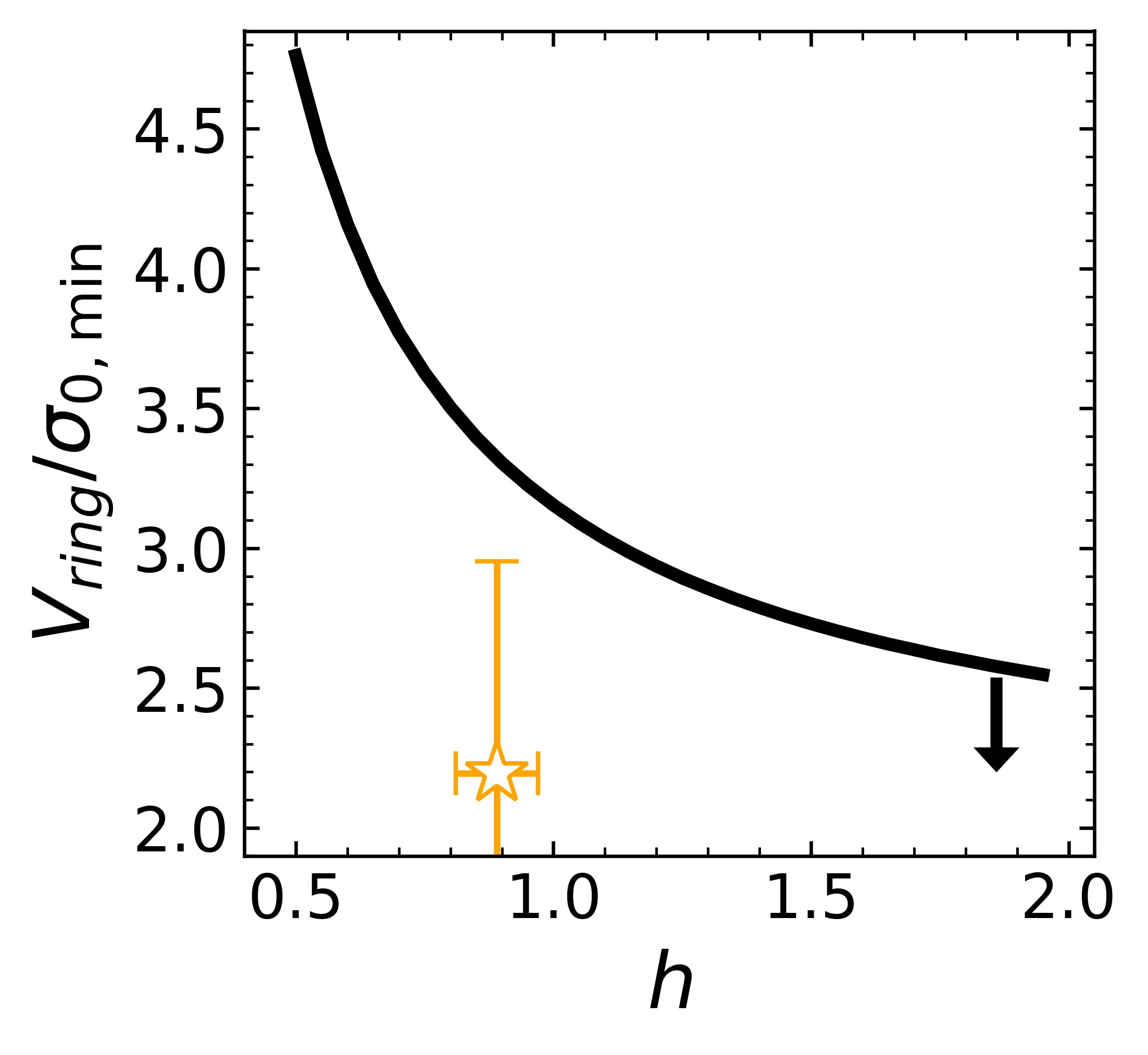}
    \caption{Maximal $V_{\mathrm{ring}}/\sigma_0$ values for which a Gaussian ring can sustain rotational equilibrium by its own self-gravity, requiring Eq.~(\ref{eq:burkert_gaussian_ring}) is non-negative everywhere. The orange star shows our bestfit values for the massive ring of zC406690 (see \S~\ref{sec:5.1general}).}
    \label{fig:GaussRing_max_v_over_sigma}
\end{figure}
Furthermore, for $r < R_{\mathrm{p}}$, pressure support will act as a stabilizing mechanism for the ring, as it provides an additional radial force-term inward. A sufficiently high velocity dispersion stabilizes the ring if the correction term is greater than the circular instability within $r_{\mathrm{circ}}$. As the correction term increases with $\sigma_0$, we can derive the minimal velocity dispersion $\sigma_{0,\mathrm{min}}$ to reach centrifugal equilibrium at all radii $r < r_{\mathrm{circ}}$. Of course, if the velocity dispersion is too high ($V/\sigma_0 \lesssim 1$), the system will no longer be rotationally supported. We examine the values of $\sigma_{0,\mathrm{min}}$ (in terms of the maximal $V_{\mathrm{circ}}/\sigma_0$) for which the rotational equilibrium criteria is satisfied (see \S~\ref{sec:3.1Gaussain Ring}) as a function of the ring's shape parameter $h$. Figure~\ref{fig:GaussRing_max_v_over_sigma} shows that for typical values of $h$ the ring is still rotationally supported ($V/\sigma_0 \lesssim 2.2-3$) and can sustain circular orbits purely due to its own pressure support. Introducing additional mass components would increase the upper limit imposed on $V_{\mathrm{circ}}/\sigma_0$.

\subsection{Rotation Curve Fitting} \label{sec:4.2RC fitting}
\begin{table*}[t]
    \centering
\resizebox{\textwidth}{!}{%
\begin{tabular}{c||cc||cc||cc}
 & \multicolumn{2}{c}{(A) Dynamical ring}& \multicolumn{2}{c}{(B) Flux ring} & \multicolumn{2}{c}{(C) Contracted halo}\\
         &2D&\texttt{dysmalpy}&2D&  \texttt{dysmalpy}& 2D&\texttt{dysmalpy}\\\hline \hline
 free parameters& 6 & 6 & 7 & 7 & 5 & 5\\ 
         $\chi^2_{red}$ &0.99&1.15&1.42&  1.54& 1.57&1.71\\
 AIC& 0& 0& 9.7& 12.9& 11.8&15.0\\ \hline
 $\log {M_{\mathrm{baryon}} / M_\odot }$& $10.99^{+0.05}_{-0.06}$& $11.02^{+0.07}_{-0.09}$& $11.05^{+0.05}_{-0.06}$& $11.03^{+0.09}_{-0.08}$& $9.91^{+0.23}_{-0.27}$&$10.25^{+0.18}_{-0.32}$\\
  $\log {M_{\mathrm{ring}} / M_\odot}^*$&$10.10^{+0.30}_{-0.34}$& $10.39^{+0.19}_{-0.27}$ &-& -& $9.91^{+0.23}_{-0.27}$&$10.25^{+0.18}_{-0.32}$\\
         $R_{\mathrm{p}} / \, \mathrm{kpc}$&$4.52^{+0.33}_{-0.34}$& $4.60^{+0.27}_{-0.34}$&$4.55^{+0.34}_{-0.34}$&  $4.54^{+0.36}_{-0.25}$& $4.57^{+0.11}_{-0.10}$&$4.44^{+0.34}_{-0.27}$\\
 $h$& $0.89^{+0.08}_{-0.09}$& $0.89^{+0.06}_{-0.09}$&$0.89^{+0.09}_{-0.09}$& $0.90^{+0.09}_{-0.07}$& $0.94^{+0.06}_{-0.07}$&$0.89^{+0.09}_{-0.08}$\\
 $B/T$& $0.87^{+0.09}_{-0.10}$& $0.76^{+0.18}_{-0.22}$& $0.78^{+0.16}_{-0.23}$& $0.88^{+0.11}_{-0.28}$& -&-\\
         $\log {M_{\mathrm{bulge}} / M_\odot}^*$&$10.93
^{+0.07}_{-0.08}$& $10.90^{+0.07}_{-0.08}$&$10.94^{+0.10}_{-0.14}$&  $10.97^{+0.11}_{-0.16}$& -&-\\
 $\log {M_{\mathrm{disk}} / M_{\odot}}^*$& -& -& $10.39^{+0.32}_{-0.46}$& $10.11^{+0.41}_{-1.02}$& -&-\\
 $R_{\mathrm{eff,disk}} / \, \mathrm{kpc}$& -& -&$4.4^{+2.1}_{-1.6}$& $6.2^{+1.4}_{-2.6}$& -&-\\
 $f_{\mathrm{DM}} (R_{\rm eff})\, ^*$ & $0.07^{+0.10}_{-0.04}$& $0.02^{+0.20
}_{-0.02}$& $0.06^{+0.08}_{-0.04}$& $0.03^{+0.18}_{-0.03}$& $0.97^{+0.01}_{-0.02}$&$0.86^{+0.07}_{-0.02}$\\
$\log {M_{\mathrm{vir}} / M_\odot}$& $10.80^{+0.90}_{-0.92}$& $9.75^{+2.72}_{-1.07}$& $10.80^{+0.85}_{-0.85}$& $10.05^{+2.48}_{-0.75}$& $11.92$&$11.92$\\
 $\alpha_{\mathrm{NFW}}$& 1& 1& 1& 1& $1.88^{+0.04}_{-0.04}$&$1.87^{+0.06}_{-0.05}$\\
 $\sigma_0 / \, \mathrm{km\, s^{-1}}$& $67^{+4}_{-5}$& $77^{+5}_{-4}$& $66^{+5}_{-5}$& $76^{+6}_{-4}$& $72^{+4}_{-4}$&$80^{+4}_{-6}$\\ \hline
 $\log {M_{\mathrm{tot}} (< 1\, \mathrm{kpc}) }$& $10.63$& $10.60$& $10.65$& $10.67$& $10.19$& $10.18$
 \end{tabular} %
 }
\caption{Best-fit values of all three models for zC406690, for the two fitting procedures. Parameters marked with an asterisk are inferred and not directly fitted for, except in the case of $f_{\rm DM}$ and $\log M_{\rm vir}$: the 2D tool uses the virial mass as a free parameter and infers $f_{\rm DM} (R_{\rm eff})$, while \texttt{dysmalpy} uses $f_{\rm DM} (R_{\rm eff})$ as a free parameter and infers $\log M_{\rm vir}$. The goodness of fit is characterized by the reduced Chi-squared test and using the Akaike Information Criterion (AIC; numbers referring to differences from the best-fitting model for each fitting method separately).} 
\label{tab:zc406690 results}
\end{table*}

We fit the 1D rotation curve along the major-axis, using a Markov Chain Monte Carlo Bayesian analysis exploring the posterior distribution of our model parameters simultaneously, using the \texttt{PYTHON} package \texttt{emcee} \citep{Foreman-Mackey2013}.  We use the major-axis RC as it is less sensitive to deviations from circular motions which are more noticeable along the minor-axis \citep{Price2021_rc41, Genzel2023}. We adopt both (i) the forward-modeling code \texttt{dysmalpy}\footnote{https://www.mpe.mpg.de/resources/IR/DYSMALPY}, implementing a full 4D reconstruction of the intrinsic space ($x_{\mathrm{gal}}, y_{\mathrm{gal}}, z_{\mathrm{gal}}, v$) and projecting it on the sky coordinated ($x_{\mathrm{sky}}, y_{\mathrm{sky}}, v$) taking into account the instrumental Line-Spread-Function (LSF) and beam smearing effects over the PSF \citep{Davies2004_dysmalpy, Davies2004b_dysmalpy, Cresci2009_dysmalpy, Davies2011, Wuyts2016, Lang2017, Price2021_rc41, Lee2025} and (ii) a 2D forward modeling tool, taking the galactic mid-plane to reconstruct the observed velocity field taking into account instrumental and beam smearing effects \citep{nestor_shachar_rc100_2023}.

The MCMC fitting procedure is run using 400 walkers with 300 iterations each after a burn-in phase of 100 iterations. Our mean acceptance rate for all models is $\approx 0.2$, which might indicate slow convergence, but encourages parameter exploration. Reviewing the walker trace plots verifies that our parameter space is well explored with $>50$ times the autocorrelation time. The bestfit values are estimated as the median of the posterior distribution, with a 1$\sigma$ uncertainty determined from the $16^{\rm th}$ and $84^{\rm th}$ percentiles.

\begin{table}[h]
    \centering
    \begin{tabular}{c c c c c c }
          & Model & DM halo & Bulge& Ring& Disk\\ \hline \hline
         (A) & Dynamical ring& NFW& \checkmark & \checkmark & -\\
 (B) & Flux ring& NFW& \checkmark & -& \checkmark \\
 (C) & Contracted halo& $\alpha$NFW& -& \checkmark & -\\\end{tabular}
    \caption{Mass components for each of the models considered. In all models the central bulge is dark with respect to the tracer.}
    \label{tab:model components}
\end{table}
\begin{table}[h]
    \centering
    \begin{tabular}{c|c c}
         parameter & initial value & prior type \\ \hline \hline
         $\log {M_{\mathrm{baryon}} /\ M_{\odot}}^{[a]}$& $11.2$ & $N \left[ 0.5 \right]$ \\
         $B/T$ & 0.5 & $U\left[ 0,1 \right]$\\
         $R_{\mathrm{p}} /\ \text{kpc}^{[b]}$& $4.6$ & $N \left[ 0.15 \right]$ \\
         $FWHM_{\mathrm{ring}} /\ \text{kpc}^{[b]}$& $5$ & $N \left[ 0.3\right]$\\
 $R_{\mathrm{eff}} /\ \text{kpc}^{[c]}$& $4.8$ &$N \left[ 2 \right]$ \\
         $\sigma_0 /\ \text{km}\ \text{s}^{-1}$ & $60$ & $U \left[0, 200\right]$ \\
         $\log {M_{\mathrm{vir}} /\ M_{\odot}}^{[d]}$& $11.9$ & $U \left[7, 14 \right]$ \\
 $\alpha_{\mathrm{NFW}}\, ^{[e]}$& 1&$U[0,3]$\\
         $c_{\mathrm{NFW}}\, ^{[f]}$& $4$ & fixed\\
         $i$ & $25^{o}$ & fixed\end{tabular}
    \caption{Parameters of the MCMC fitting and their priors. $U \left[ a, b \right]$ denotes a uniform distribution between a and b and $N \left[ s \right]$ a Normal distribution with a standard deviation of $s$. [a] SED fitting from \citealt{ForsterSchreiber2018} and gas fractions from \citealt{Tacconi2018}. [b] Fitting result of a Gaussian ring for the H$\alpha$ flux. [c] H$\alpha$ effective radius from \citealt{ForsterSchreiber2018}.  [d] stellar-mass halo-mass relations from \citealt{Moster2018}. [e] Fitted for only in model C, otherwise fixed at $\alpha_{\rm NFW} = 1$. [f] cosmological halo relations from \citealt{Dutton2014}.}
    \label{tab:parameter priors}
\end{table}

We use ancillary and simulation-based data to determine the prior probabilities and fix some parameters (see Table~(\ref{tab:parameter priors})). The inclination is based on previous 2D optical \emph{HST} imaging \citep{Lang2014, Tacchella2015}, and the initial baryon mass estimated as a sum of the stellar mass $M_\star = 4.14 \times 10^{10}\, \rm M_\odot$, and assuming cold-to-stellar mass ratio dependent on the properties of the galaxy $\mu_{\rm gas} = M_\star / M_{\rm gas} \approx 2.5$. \citep{ForsterSchreiber2018,Tacconi2018}. This gives $\log {\left( M_{\mathrm{baryon, init}}/M_\odot \right)} = 11.2$ which is then free to vary with a Gaussian prior with a 0.5dex standard deviation. A fraction of the total baryons is associated with a central bulge, defined via the bulge-to-total mass ratio $B/T = M_{\rm bulge} / M_{\rm baryon}$, which is uniformly sampled between [0, 1] with an initial value of $B/T \,_{\rm init} = 0.5$. When considering a ring (models A and B), we take the initial values of the parameters determining the shape of the Gaussian ring, $R_{\mathrm{p}}$ and $h$, by fitting the $\mathrm{H\alpha}$ flux. We find $R_{\rm peak} = 4.6 \pm 0.15\, \rm kpc$ and $\rm FWHM_{\rm ring} = 5.0 \pm 0.3\, \rm kpc$ (corresponding to $h = 0.92 \pm 0.064$) for the initial values with a Gaussian prior with a standard deviation of the uncertainties. When considering a disk (model B), the initial effective radius is based on the H$\alpha$ effective radius from \cite{ForsterSchreiber2018} with a Gaussian prior, and the intrinsic thickness of the disk is fixed to $q_0 = 0.25$, a typical value for thick disks at these redshifts \citep{VanDerWel2014a}.

The virial mass, $M_{\mathrm{vir}}$, is initially set to match cosmological stellar-halo mass relations \citep[e.g.][]{Moster2018} and allowed to vary with a uniform prior distribution (\texttt{dysmalpy} fits instead for $f_{\rm DM}(R_{\rm eff})$ directly, which is degenerate with $M_{\rm vir}$). The concentration of the NFW profile is fixed to $c_{\mathrm{NFW}}=4$, based on redshift dependent scaling relation \citep{dutton_cold_2014}. Finally, the velocity dispersion $\sigma_0$ is fitted for with a uniform prior and an initial value typical of high redshift disks, $\sigma_0 = 60 \, \rm km/s$ \citep{Ubler2019}. It is important to note that considering a radially decreasing dispersion instead of a constant value will result in lowering the dark matter mass and lower $f_{\rm DM}$, as they both affect the outer part of the RC. The reduced pressure support due to the dispersion gradient will be counteracted by lower amounts of dark matter, in order to reach similar values of $v_{\rm rot}$. In total, our free parameters are $M_{\rm baryon}, B/T, R_{\rm p}, FWHM_{\rm ring}, M_{\rm vir}$ (or $f_{\rm DM}(R_{\rm eff})$) and $\sigma_0$.

With these assumptions in place, we consider three different mass models representing three different scenarios (see Table~\ref{tab:model components}). Model (A) is the case of a dynamical Gaussian ring with a constant M/L, a central bulge and an NFW halo; model (B) consists of an exponential disk, a central bulge and an NFW halo, with the addition of a Gaussian ``flux ring" affecting flux weighting only; and model (C) for the case where the bulge is substituted with a contracted NFW halo ($\alpha >1)$, consisting of a dynamical Gaussian ring and a $\alpha$NFW halo. Previous studies have considered explicit disk+bulge configurations in a NFW halo, but have not directly modeled the effect of a ring \citep{Genzel2017, Price2021_rc41, nestor_shachar_rc100_2023}. Moreover, deviations from NFW might be an additional method to creating a compact central distribution (for $\alpha_{\rm NFW} > 1$), required by the previous modeling but lacking a detection. The models we consider will provide a better fit to this galaxy, more suited to its unique characteristics and given the constraints from various observations.

\section{Results} \label{sec:5.results}
\subsection{Mass Decomposition}\label{sec:5.1general}
\begin{figure*}[t]
    \centering
    \includegraphics[width=0.7 \linewidth]{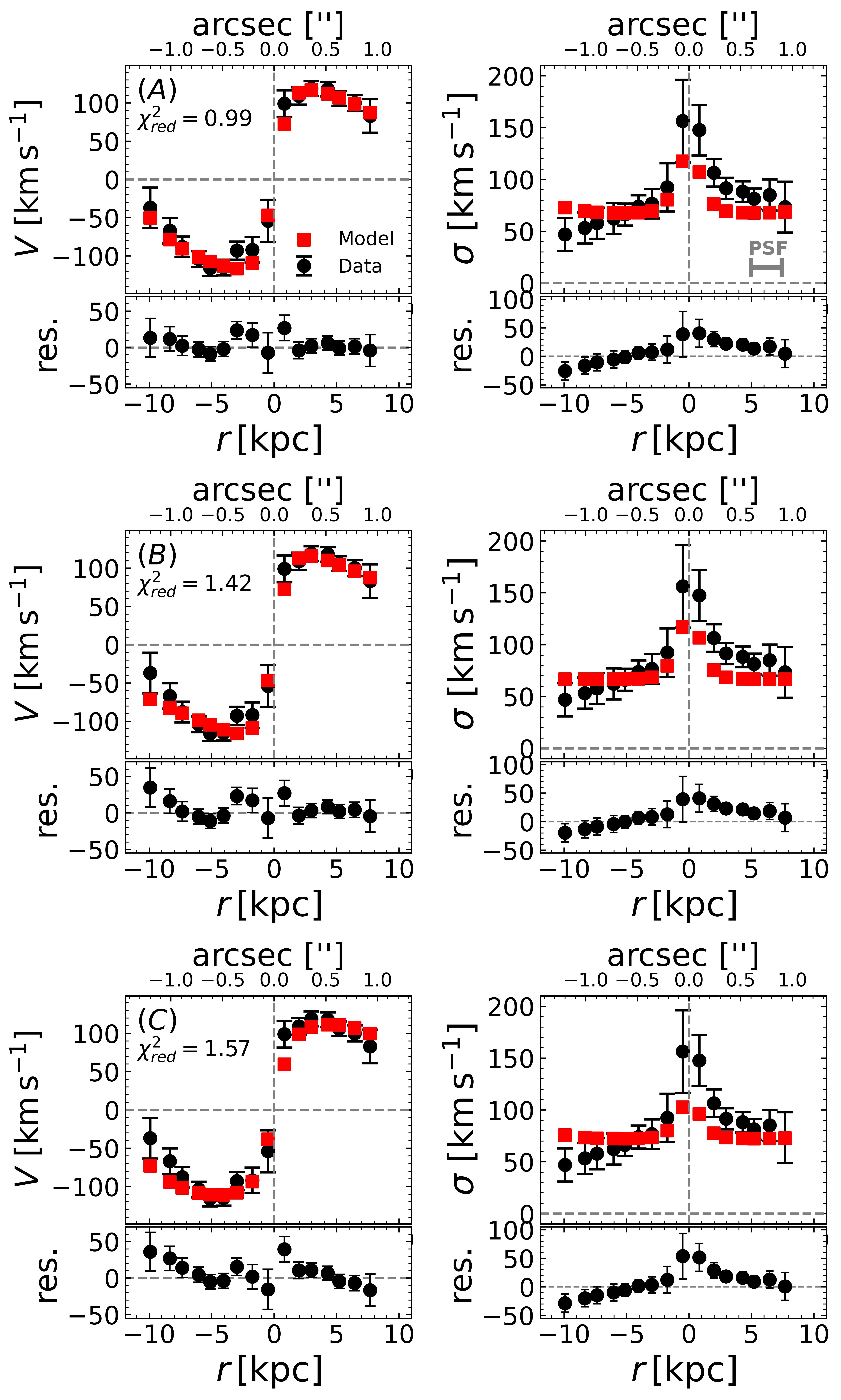}
    \caption{Best-fit rotation curve and velocity dispersion for the 2D modeling tool, along with the residuals, for the three models listed in Table~(\ref{tab:model components}). Black and red dots mark the data and model, respectively, and the residuals are $\Delta V = V_{\mathrm{data}} - V_{\mathrm{model}}$. The PSF FWHM is given by the grey horizontal bar. The outer part of the RC contains the most important information in differentiating the models.}
    \label{fig:RC_bestfit}
\end{figure*}
The kinematics of zC406690 is best described using a dynamical massive ring (A), capturing the nature of a rapid increase in velocity and a short flattening, followed by a rapid (super-)Keplerian decline (see Table~(\ref{tab:zc406690 results})). Both a flux-emitting ring (B) and a contracted DM halo (C) provid a worse fit missing some of these features. We find the total mass in the ring to be $M_{\mathrm{ring}} = 1.2-2.5 \times 10^{10}\, \mathrm{M_\odot}$ with a peak radius of $R_{\mathrm{peak}} = 4.5\, \mathrm{kpc}$ and a shape parameter $h = 0.9$, close to our fiducial value. Such a shape parameter implies an effective radius of $R_{\mathrm{eff}} = 5.7\, \mathrm{kpc}$, and does not impose strict limitation on the rotational stability of the ring, i.e., requiring a minimal B/T $>0.2$.  We estimate the goodness-of-fit using both the $\chi^2 _{red}$ test and Akaike's Information Criteria (AIC), both suggesting the massive ring model is superior. The AIC values indicate the other models have a probability of only $0.1\% - 0.5 \%$ to better describe the data over model (A). The fit results using \texttt{dysmalpy} are given in Appendix~\ref{app:RC_fitting_dysmalpy}, and are in excellent agreement.

The rapid decline in the rotation curve implies a baryon dominated system, for which we indeed find very low DM fractions $f_{\mathrm{DM}} < 0.07$. These values imply that close to $40\%$ of the total baryonic mass of the galaxy is found in the innermost $2$ kpc, with an enclosed mass of $\log {M \left( < 1\, \mathrm{kpc} \right) / M_\odot} \approx 10.6$. The only exception is model (C), where we replace the bulge with an $\alpha$NFW DM halo, to allow for the contraction of the halo to create such a high mass concentration. We find that the galaxy then becomes extremely DM-dominated with $f_{\mathrm{DM}} \approx 0.9$ and an inner density slope of $\alpha_{\mathrm{NFW}} = 1.9$. However, even though this builds up a large central mass concentration, $\log {M \left( < 1\, \mathrm{kpc} \right) / M_\odot} = 10.2$, due to the nature of the NFW profile the resulting rotation curve does not drop as rapidly as the other models. 

The bulge-to-total ratios are all rather high (models A \& B), and all point to a central bulge with a mass of $\log{\left( M_{\mathrm{bulge}} / M_\odot \right)} = 8-9 \times 10^{10}\, \mathrm{M_\odot}$, similar to values found in previous fitting \citep{Genzel2020, nestor_shachar_rc100_2023}. Kinematically, a large bulge is required to both create a steep inner velocity gradient and a rapid drop of the outer RC. The large velocity gradients in the center also create the ``bump" in the observed velocity dispersion, as the emission line is broadened over the beam PSF. This raises some concern, as such a stellar component is not seen in any of the available imaging, even at rest-frame 1.4 $\mathrm{\mu m}$ which is less susceptible to dust extinction (see Figure~\ref{fig:zC406690_flux_maps}). However, a very compact and dense dust distribution can attenuate the flux enough to hide such a bulge. We discuss this possibility further in the next section. 

\begin{figure}[ht]
    \centering
    \includegraphics[width=1\linewidth]{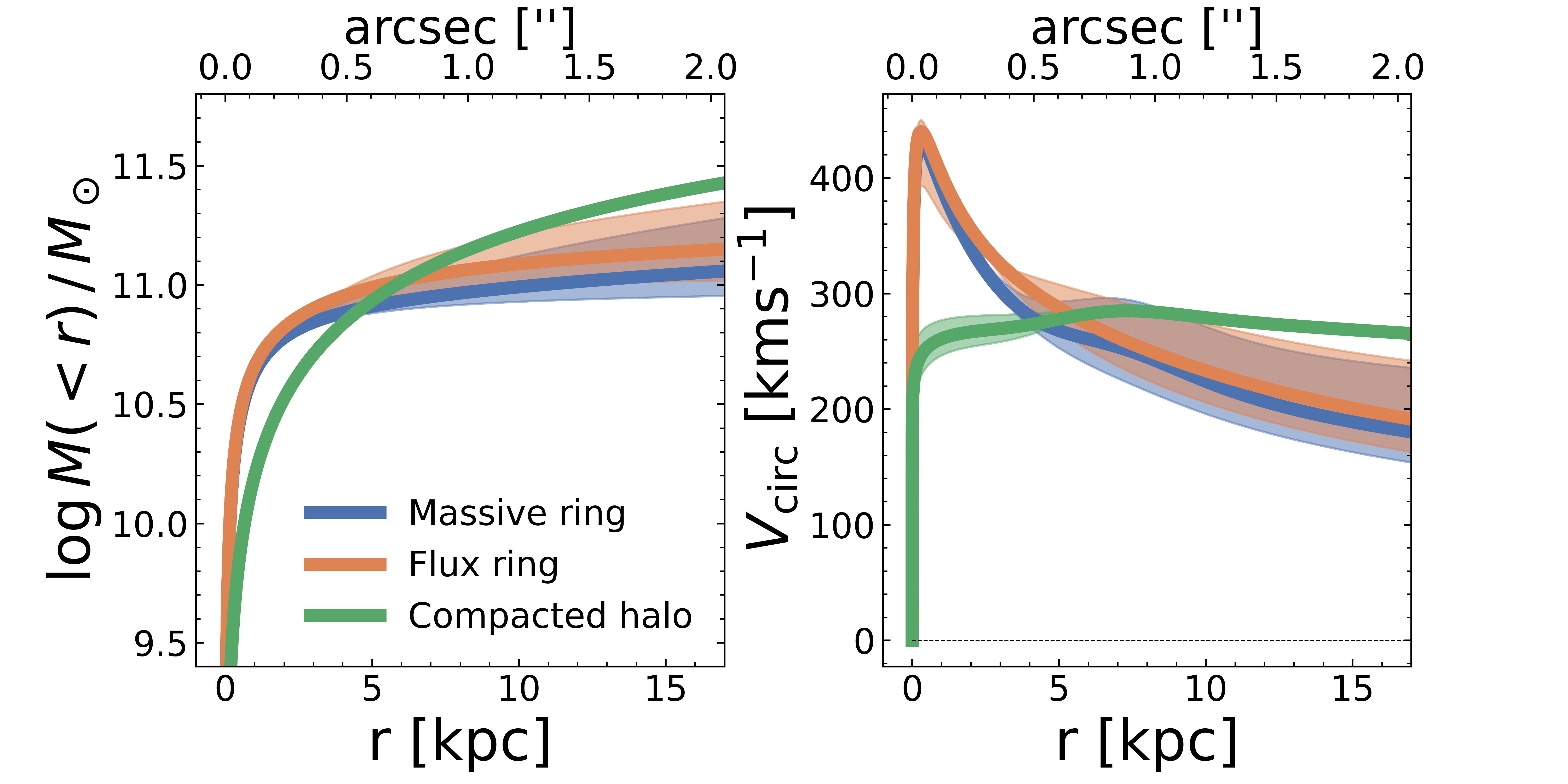}
    \caption{Enclosed total mass of baryons and dark matter (left) and the circular velocity profiles (right) for the three best-fit models. Shaded regions show the $1 \sigma$ uncertainties of each model.}
    \label{fig:enclosed_mass_profiles}
    \end{figure}

\begin{figure*}[t]
    \centering
    \includegraphics[width=15
    cm]{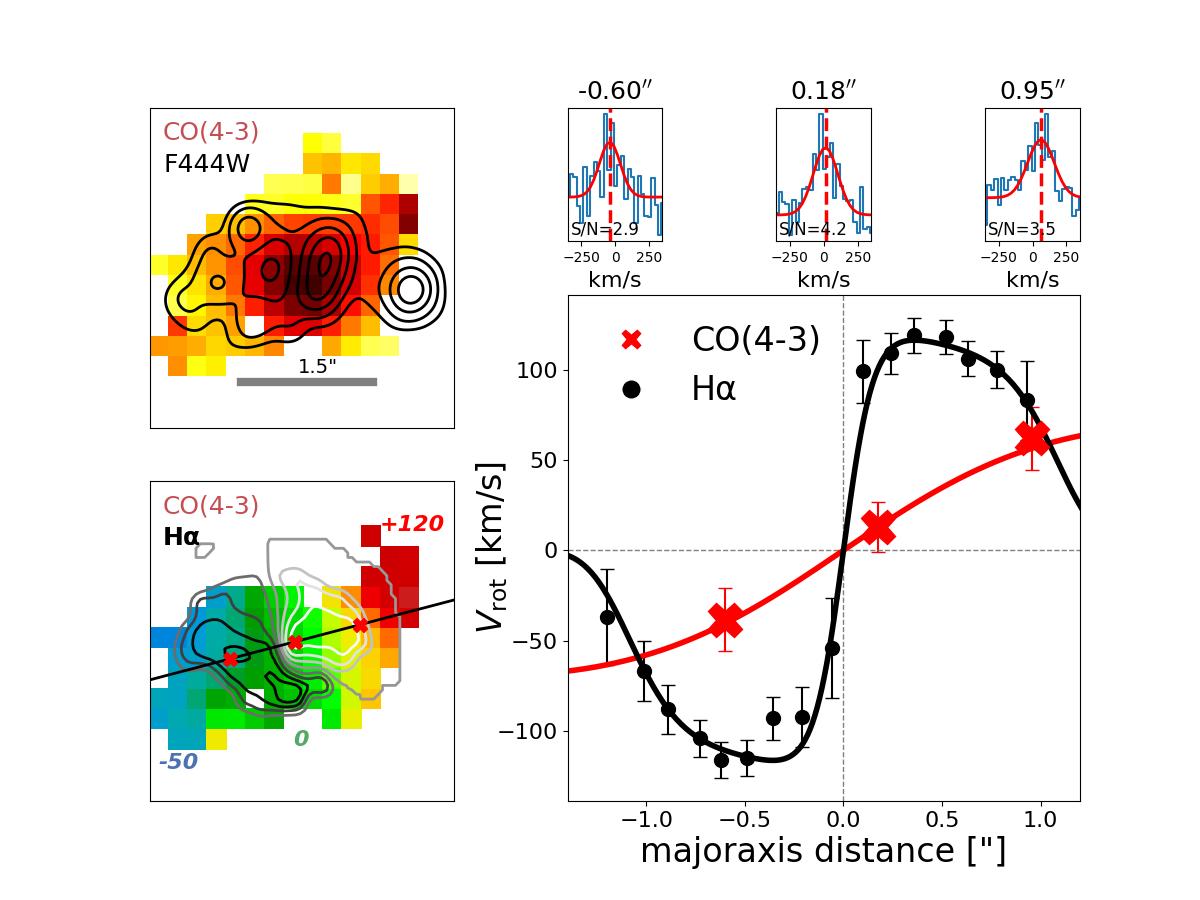}
    \caption{CO(4-3) line emission of zC406690 from ALMA (synthesized beam PSF 1.5"). Top left: CO(4-3) flux overlaid on contours from \emph{JWST} F444W (PSF 0.15"), showing the cold gas peak is between the western clump and the kinematic center. Bottom left: CO(4-3) velocity map obtained from Gaussian line fitting of each individual spaxel, overlaid on the H$\alpha$ iso-velocity contours from adaptive-optics \emph{VLT}/SINFONI \citep{ForsterSchreiber2018}. The black line shows the $H\alpha$ kinematic majoraxis, and the red crosses mark independent data points extracted for the CO(4-3) rotation curve. Top right: Line profiles of the CO(4-3) data points along the majoraxis, taking circular apertures of radius FWHM/2 and the best Gaussian fit. Right: Rotation curves of H$\alpha$ (black dots) and CO(4-3) (red crosses), the solid black line shows the bestfit Dynamical ring model (A) and the red solid line shows the same mass model mock-observed at the resolution of the ALMA observation.}
    \label{fig:CO rotation curve}
\end{figure*}
Figure ~\ref{fig:enclosed_mass_profiles} shows the cumulative mass and circular velocities for each model, emphasizing the similarities in the intrinsic total mass budget. The differences are most notable in the central few kpc where the velocity peaks sharply, and at the drop induced by pressure support at around $10$ kpc, where this effect becomes important. While statistical analysis favors the Dynamical ring (A), the subtle difference require further observations reaching high S/N in the outer galaxy ($\gtrsim 1^{\prime \prime}$), or have a high spectral resolution reaching $\sim 20\, \mathrm{kms^{-1}}$ ($R = 15\text{,}000$). While models (A) and (B) are more similar, the compacted halo scenario (C) does not reach the high velocity peak close to the center and intrinsically continues to accumulate mass in a region where the other models flatten out. However, the strong pressure support term following Eq.~(\ref{eq:burkert_gaussian_ring}) ultimately dominates $V_{\mathrm{rot}}$ in the outer part of the galaxy, making this region even more important in determining the intrinsic density profile of this galaxy.

\subsection{CO(4-3) emission line kinematics}\label{sec:5.2CO_kinematics}
\begin{figure}[ht]
    \centering
    \includegraphics[width=\columnwidth]{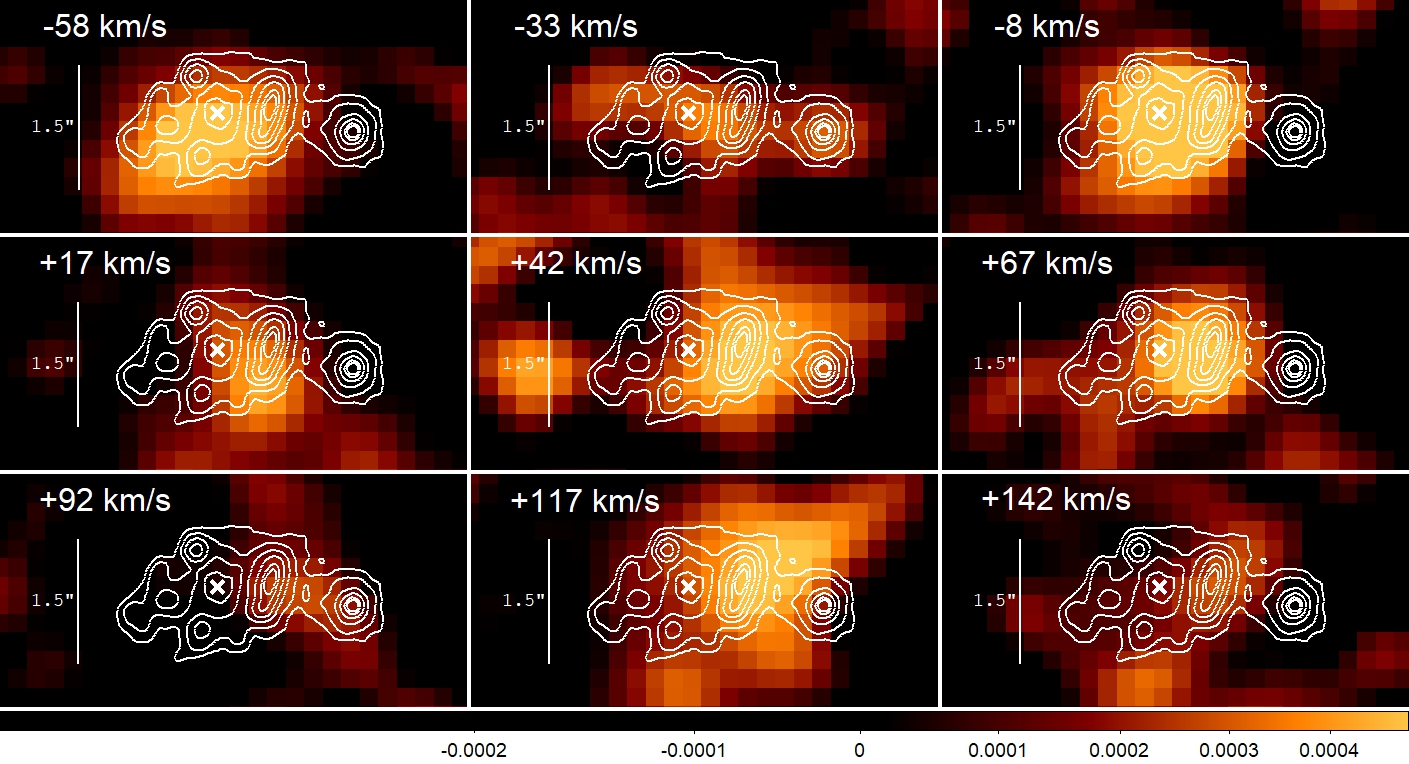}
    \caption{Channel maps of the CO(4-3) emission line for zC406690 (PSF $1.5^{\prime \prime}$), over contours tracing NIRCam F444W imaging. The channels correspond to velocities ranging from $-58\, \mathrm{km/s}$ (top-left) to $142\, \mathrm{km/s}$ (bottom-right) with $\Delta v = 25\, \mathrm{km/s}$, colored by the flux in $\rm Jy/beam$. The H$\alpha$ kinematic center is marked with a white cross.} 
    \label{fig:CO_channel_maps}
\end{figure}

Archival ALMA band 4 observations have detected a marginally resolved CO(4-3) line emission originating from the galaxy, with which we could only obtain a few kinematic data points along the major-axis. To compare to the H$\alpha$ emission, we calibrate the CO position first using \textit{JWST}/NIRCam, as both are registered to GAIA DR3 astrometry, and then align the \emph{VLT}/SINFONI moment-0 map to match the clump positions. Figure~\ref{fig:CO rotation curve} (top left panel) displays the CO(4-3) emission map on top of the contours of the NIRCam/F444W filter, indicating that the CO(4-3) peaks between the center of the ring and the western clump.

By performing individual Gaussian fitting for each spaxel, we construct a CO(4-3) velocity map that reveals a clear rotation pattern with a very similar PA as the H$\alpha$ (see Figure~\ref{fig:CO rotation curve}, bottom left panel). Given the channel width of $\Delta v = 25\, \mathrm{km\, s^{-1}}$, the emission in each channel can clearly be seen moving along the PA from the blue to the red side (see Figure~\ref{fig:CO_channel_maps}). We conclude that part of the CO emission must be originating from the ring and/or clumps along the ring, having similar rotation pattern, otherwise no rotation would be observed. Such agreements between H$\alpha$ and CO are known for SFGs at such redshifts \citep{Ubler2019, girard_systematic_2021}.

We construct the CO rotation curve from the few data points we could reliably extract, given the large PSF. We position circular apertures of $0.5\times$PSF along the major-axis that maximizes S/N and fit a Gaussian profile to determine the centroid velocity (given by the red dots on Figure \ref{fig:CO rotation curve}). The resulting rotation curve obviously cannot be fitted for directly, as our model has more degrees of freedom than data points, but we test whether our mass decomposition (see \S~\ref{sec:5.1general}) is a good fit. We create a mock model using our best-fit values for the Dynamical ring model (A), with the degraded resolution of the ALMA observation. The resulting RC (red line) is an excellent fit, matching the molecular gas kinematics very well. The shape is of course very different, as large beam-smearing effects create a ``smooth" version of the detailed H$\alpha$ rotation curve.

\subsection{Hiding the Bulge - Dust and molecular gas}\label{sec:5.3.Bulge_exntinction}
But where is the bulge? Could it be hidden by dust extinction? There is no detection of any ionized gas or stellar emission from the central region, even at the longest wavelength available NIRCam/F444W (rest-frame J-band). SFGs that have previously lacked a central stellar component with \emph{HST} imaging and were highly clumpy and irregular, were later shown to have compact bulges \citep{ferreira_jwst_2023}. However, large amounts of dust could attenuate the emission coming from the center to fall below the \emph{JWST} sensitivity. Using the observations taken with \emph{ALMA} we can test if there is enough dust to allow for such high extinction of the bulge. We consider the V- and J-bands, roughly matching NIRCam's F150W and F444W filters at this redshift.

We derive the total molecular mass from the integrated CO(4-3) line over the entire galaxy by placing a circular aperture with radius $1^{\prime \prime}$ over the center of the galaxy, and fitting a single-Gaussian for the line intensity finding $F_\nu = 167 \pm 23\, \mathrm{mJy\, km\, s^{-1}}$ and a linewidth of $\Delta V = 240 \pm 39\, \mathrm{km s^{-1}}$. We convert it to mass following \cite{Tacconi2018}:
\begin{equation}
\begin{split}  
    M_{\mathrm{mol,gas}} &= 1.58 \times 10^{9} M_{\odot} \left( \frac{F_{CO43}}{\mathrm{Jy\, km\, s^{-1}}} \right) \left( 1+z \right)^{-3} \\
    & \times \left( \frac{\lambda_{obs, 43}}{\mathrm{mm}} \right)^2 \left( \frac{D_L}{\mathrm{Gpc}} \right)^2 R_{41}\, \chi (Z) \\
    & = 7.1 \pm 1.0 \times 10^{10}\, \mathrm{M_{\odot}}
\end{split}
\end{equation}
with $D_L = 17.64\, \mathrm{Gpc}$, and standard conversion factor for the CO($4 \rightarrow 3$) emission ($R_{41} = 2.4$, $\alpha_{\mathrm{CO}} = 3.6$). We use the metallicity function $\chi(Z) = 10^{-1.27 \times \left( Z-8.67 \right)}$ from \cite{Genzel2012}, for a metallicity of $Z = 12 + \log{(O/H)} = 8.38$ \citep[][and see table \ref{tab:zc406690_properties}]{genzel_sins_2011}. Together with the stellar mass this yields a cold gas fraction of $\mu_{\mathrm{gas}} = 1.65 \pm 0.34$, typical for a MS galaxy at this redshift \citep{Tacconi2018}. Based on the SFRs of the brightest clumps, they contain a molecular gas mass of $M_{\mathrm{mol,clumps}} = 3.3 \times 10^{10}\, M_{\odot}$, roughly half the total molecular gas of the galaxy (assuming they follow standard Kennicutt-Schmidt relations, and see Table~\ref{tab:zc406690_properties}). Therefore, we can place an upper limit on the amount of gas, and the dust associated to it, available to play a role in the attenuation of the bulge, $M_{\mathrm{mol,bulge}} \lesssim 4 \times 10^{10}\, M_{\odot}$. The strong winds detected at the location of the brightest clumps may imply significant gas depletion, and thus lower $M_{\rm mol,clumps}$ than derived from the Kennicutt-Schmidt relation, increasing the available gas. On the other hand, some fraction of the molecular gas may also be associated with the other, fainter, clumps along the ring. We therefore adopt the upper limit above in what follows, keeping in mind there are significant uncertainties. 

The dust-to-gas ratio of the galaxy is important to correctly determine the attenuation. We detect a continuum emission in the $142$ GHz ALMA band 4 (rest-frame $453$ GHz), from which we can derive the total dust mass $M_{\mathrm{dust}}$ \citep[following][]{Magnelli2020}:
\begin{equation}
\begin{split}  
    M_{\text{dust}} &= \frac{5.03 \times 10^{-31} \times S_{\mathrm{\nu}} \times D_{\mathrm{L}}^2}{\left( 1+z \right)^4 \times B_{\nu_{\text{obs}}} \left( T_{\text{obs}} \right) \times \kappa_{\nu_0}} \left( \frac{\nu_0}{\nu_{\text{rest}}} \right)^\beta \\
    & = 2.97 \pm 0.66  \times 10^8\, \mathrm{M_\odot}
\end{split}
\end{equation}
where the observed flux density is $S_{\mathrm{\nu}} = 44.0 \pm 9.5 \, \mathrm{\mu Jy}$, $B_{\nu_{\text{obs}}} \left( T_{\text{obs}} \right)$ is the Planck function in $\mathrm{Jy\, sr^{-1}}$ for an observed dust temperature of $T_{\text{obs}} = T_{\text{rest}} / (1+z)$ (taking a rest-temperature of $25\, \text{K}$), $D_{\text{L}}$ is the luminosity distance in meters, $\beta = 1.8$ is the dust emissivity, $\kappa_{\nu_0}=0.0431\, \mathrm{m^2\, kg^{-1}}$ is the photon cross-section to mass ratio of dust at rest-frequency $\nu_o = 352.6\, \mathrm{GHz}$. We assume corrections due to CMB photons typically negligible at this redshift and an optically thin emission. Given the total molecular gas mass, this yields a dust-to-gas ratio (DGR) of $\delta_{\rm d} = M_{\mathrm{dust}}/M_{\mathrm{gas}} = 0.0068 \pm 0.0017$, slightly lower than the standard MW value $0.01$ \citep[see][for example]{Sandstrom2013}. The metallicity inferred from the DGR based on scaling relations is $12+\log {\mathrm{O/H}} = 8.23\pm 0.18$ \citep{Tacconi2018}, in good agreement with the estimate based on the ionized gas  (see Table \ref{tab:zc406690_properties}).

For a hydrogen column density $\mathrm{N_H}$, the associated V-band attenuation $A_{\rm v}$ can be computed as $A_{\mathrm{V}} = 5.3\ \mathrm{N_{H_2}} \left( \frac{\delta_{\rm d}}{0.01} \right) \times 10^{-22} \mathrm{cm^2}$ \citep{Draine2011}. At $z=2$, neutral hydrogen is typically negligible on galactic scales, and it can be assumed that $\mathrm{N_H} \approx \mathrm{N_{H_2}}$ \citep{Tacconi2020}. Denoting the fraction of molecular gas covering the bulge as $f_{\mathrm{mol,bulge}} = M_{\mathrm{mol, bulge}} / M_{\mathrm{mol}}$, the Hydrogen column density within an aperture of $2$ kpc (matching the assumed bulge size) is $\mathrm{N_{H_2}} = 1.4\times 10^{24} f_{\mathrm{mol.bulge}}\, \mathrm{cm^{-2}}$. Correcting for the observed dust-to-gas ratio, we obtain a V-band attenuation of:
\begin{equation}
    A_{\mathrm{V}} =315 f_{\mathrm{mol,bulge}}\ \ \ .
\end{equation}
The \emph{JWST}/NIRCam imaging with F150W, F277W and F444W corresponds to the g-, i-, and J-band respectively, for which $A_{\rm g}/A_{\rm V}=1.24$, $A_{i}/A_{\rm V}=0.62$ and $A_{\rm J}/A_{\rm V}=0.27$ \citep{Draine2003}. We consider two cases for the absorbing dust: (i) assuming the dust is distributed between the observer and the bulge (``screen") attenuating the flux as $\propto e^{-\tau_\lambda}$, and (ii) assuming the dust is well mixed with the stellar population (``mixed") attenuating the flux $\propto \left( 1 - e^{-\tau_\lambda} \right)/\tau_\lambda$, with the optical depth being $\tau_\lambda = A_\lambda / 1.038$. 

\begin{figure*}[ht]
    \centering
    \includegraphics[width=0.8\linewidth]{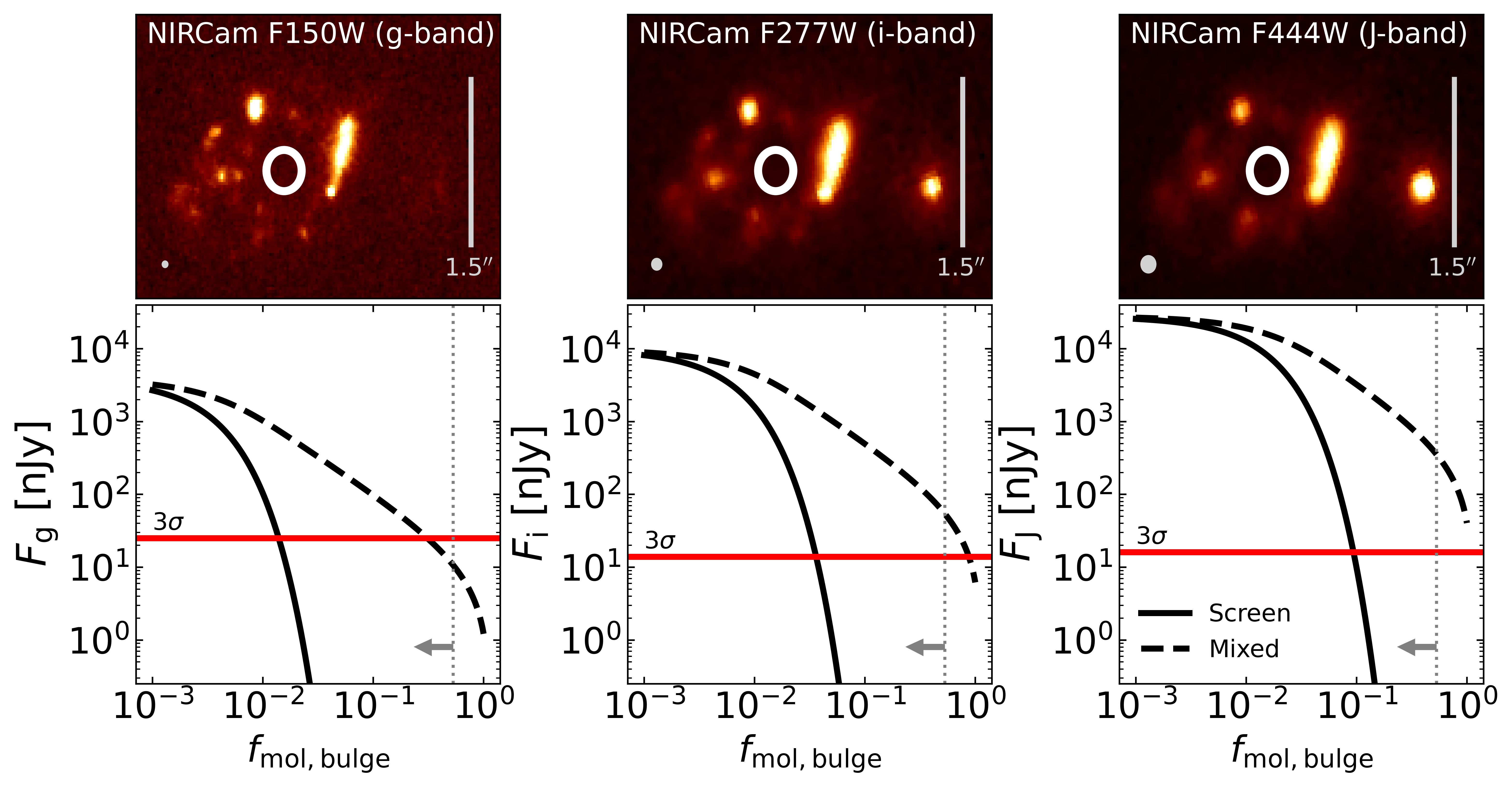}
    \caption{Expected dust-attenuated flux density from a $8 \times 10^{10} \, \mathrm{M_{\odot}}$ bulge as a function of the fraction of gas in the innermost $2\, \mathrm{kpc}$ (white circle), $M_{\mathrm{mol,bulge}} = f_{\mathrm{mol, bulge}}\, 7.1 \times 10^{10} \, \mathrm{M_{\odot}}$. Top row: \emph{JWST}/NIRCam imaging in F150W, F277W and F444W corresponding to rest-frame g-, i- and J-bands at this redshift. The grey circle shows the PSF FWHM, and the 1-sigma RMS is $8.3$, $4.6$ and $5.4 \, \rm{nJy}$. in each band respectively, (following \citealt{Casey2023}). Bottom row: expected flux density for a mixed (dashed line) and screen (solid line) dust extinction models. The red lines mark a $\rm {S/N} = 3$ signal. The vertical dotted line and the arrow mark the upper limit on available gas to be over the bulge, set by the gas locked in the clumps. Roughly $7\times 10^9\, \mathrm{M_{\odot}}$ ($10$\%) of the molecular gas has to act as a screen between the bulge and the observer in order for the flux to drop below the detection limit.}
    \label{fig:bulge dust absorption}
\end{figure*}

Figure~(\ref{fig:bulge dust absorption}) shows the expected observed flux of the bulge as a function of the fraction of molecular gas in the bulge $f_{\mathrm{mol,bulge}}$, for mixed (dashed line) and screen (solid line) attenuation models. For the emission from the bulge, we take its stellar content to be $M_{\rm bulge,\star} = M_{\rm bulge}-f_{\rm mol,bulge} \times M_{\rm mol, gas}$, as the molecular gas taking part in the attenuation also has a dynamical contribution. We take $M/L$ ratios of $M/L_{\rm g} = 1.1$, $M/L_{i} = 0.8$, and $M/L_{\rm J} = 0.45$, based on $\left( g-i \right)$ color relations calibrated in SFGs up to $z = 1.5$ \citep{Zibetti2009, LopezSanjuan2019}.

The flux can drop bellow a $3\sigma$ (dotted red line) detection in all filters only when at least $\gtrsim 10$\% of the total molecular gas, i.e. $\gtrsim 7.1\times 10^{9}\, M_{\odot}$, is acting as a screen in front of the bulge. This is consistent with the upper limit we have set earlier for the available cold gas. If so, why is the emission map in Figure~(\ref{fig:flux_ring_effects}) not centered on the kinematic center, but is somewhere between the center and the western clump? The emission that would  originate from the attenuating mass would be to weak to be directly resolved, and large cold gas reservoirs in the clumps are most likely the dominant contribution. The current ALMA observations are x5 shorter than what would be needed to achieve a detection. Future observations at higher resolution and sensitivity could directly resolve this central emission.

\section{Discussion} \label{sec:6.discussion}
zC406690 has global properties typical for a near main sequence star-forming galaxy at $z=2.196$, with a stellar mass of $M_\star=4\times 10^{10}\, \mathrm{M_\odot}$ and $\mathrm{SFR}=300\, \mathrm{M_\odot\, yr^{-1}}$. We calculate its gas fraction $M_{\rm gas} / M_\star = 1.65 \pm 0.34$ and its metallicity $12 + \log{\left( \rm {O/H} \right)}=8.23\pm 0.15$ based on the observations of the molecular gas, which are in good agreement with estimates based on the ionized gas and consistent, although slightly lower, than suggested by $M_\star-\rm Z$ relations \citep{Wuyts2016a}. Both the ionized and molecular gas kinematics suggest an ordered rotation pattern with similar velocities and PA. However, its morphology shows a striking giant ring persistent in rest-frame UV, optical and NIR, suggesting the stellar population follows the same distribution, as shown in Figure~(\ref{fig:zC406690_flux_maps}). We find that such case, where the surface density traces a Gaussian ring, best captures the kinematic features such as the super-Keplerian drop in the outer RC. This suggests that $M_{\mathrm{ring}} = 2.5 \pm 1.5 \times 10^{10}\, \mathrm{M_\odot}$ is distributed in the form an actual mass ring with a diameter of $9\, \mathrm{kpc}$.

\textbf{\textit{Are long lived rings feasible?}} Simulations suggest accretion of gas onto the disk could inject enough angular momentum to sustain the formation of large-scale massive rings \citep{dekel_origin_2020}. Longevity is defined when the accretion timescale is shorter than the typical infall time, and is achieved when the ratio of cold gas in the ring to the dynamical mass enclosed within the ring is $M_{\mathrm{gas}} / M_{\mathrm{dyn}} \lesssim 0.3$. Our modeling (Model A) suggests a dynamical mass of $M_{\mathrm{dyn}}( < R_{\mathrm{p}}) = 1.1 \times 10^{11}\, M_{\odot}$. Taking the cold gas estimate from \cite{genzel_sins_2011} we find this ratio to be $0.3$. However, this value only comes from the brightest clumps, and there could me more gas in the ring. If we take the limiting case, where the ring contains all of the molecular gas except for the amount required to attenuate the bulge (see \S~\ref{sec:5.3.Bulge_exntinction}), e.g. $M_{\rm gas} = 6.4 \times 10^{10}\, \rm{M_\odot}$, this ratio increases by almost a factor of 2. 
The outflow signatures detected in the bright clumps might suggest they are more gas depleted than suggest by KS relations, but conversely, fainter clumps along the ring could contain additional cold gas. Exact determination of gas-to-dynamical mass ratio is therefore highly uncertain without better resolved CO emission, and can only be roughly estimated. If indeed the bright clumps are gas depleted, star formation in the clumps would quench rapidly in just under $\lesssim 100\, \mathrm{Myr}$ unless fed by streams by comparable rates of $\sim 10-100\, \mathrm{M_\odot\, yr^{-1}}$, which would also act as a source of angular momentum to the ring.

\textbf{\textit{Companion interaction.}} The \emph{JWST} imaging reveals a possible interacting satellite, plausible given our redshift estimate $z_{\mathrm{phot}} = 2.14^{+0.32}_{-0.42}$. It is clearly seen only in the F277W \& F444W filter. If indeed it is the same redshift as zC406690 it must be a quenched system with substantial dust extinction. It is part of the ALMA FOV but there is no detection of either dust continuum or CO(4-3) emission, possibly due to the shallow signal. Without spectroscopy, there is very little that can be said about any interaction between the companion and zC406690. Speculatively, it might have pierced through the center of zC406690, create outward shockwaves driving material outwards. We consider such a case to be unlikely, as it would have been observed in the kinematics of the ionized gas and impact the velocity map, and any such interaction would likely strip the companion of its gas. Yet, zC406690 has a low gas content making this scenario less likely. With this data at hand, we conclude it is unlikely to be a satellite of zC406690. 

\textbf{\textit{Where is the central bulge?}} Our kinematic modeling suggests $M_{\mathrm{bulge}} = 8 \times 10^{10}\, \mathrm{M_{\odot}}$, but it has no trace in any of the observations. In \S~\ref{sec:5.3.Bulge_exntinction} we explore the possibility that it is highly extincted by dust, and show that the current observation cannot rule out such scenario. In fact, $10\%$ of the total cold gas, and dust associated with it, are enough to absorb all of the light emitted from the central bulge, leaving a dark patch. We can place direct observational predictions for the flux emitted over the central $2\, \mathrm{kpc}$ ($0.24^{\prime \prime}$). If half of the available molecular gas (see \S~\ref{sec:5.3.Bulge_exntinction}) is found in the bulge, it could be observed at a S/N$=5$ in an integration time of 10.5 hours using \emph{ALMA} band 4, at a resolution of $0.25^{\prime \prime}$. 

\section{Summary}\label{sec:summary}
In this paper we revisited the case of the SFG zC406690 ($z=2.196$) in light of recent \emph{JWST}/NIRCam and archival ALMA data. Combined with previous \emph{HST} and high-resolution integral field spectroscopy from VLT/SINFONI-AO, we have shown that a large-scale clumpy ring is seen in multiple wavelengths from rest-frame UV to NIR, with, so far, no detectable emission coming from the center (see Figure~(\ref{fig:zC406690_flux_maps})). However, a smooth rotation pattern is seen in high-resolution for the ionized gas, and in low resolution for the molecular phase, both showing similar velocity maps (see Figure~(\ref{fig:CO rotation curve})). The rotation curve implies the presence of a central mass.

We introduce a mass profile to describe a Gaussian ring surface density. A razor-thin profile provides an excellent approximation to a full 3D distribution, and we derive its key properties and the circular velocity associated with it (see also Appendix~(\ref{app:GaussianRing})). Due to its geometry, there is always a region interior to the ring that cannot sustain rotational equilibrium. This can be solved by either adding a second mass component to deepen to gravitational potential, or by corrections to the rotational velocity due to pressure support (``asymmetric drift"). As velocity dispersion increases with redshift \citep{ForsterSchreiber2018, Ubler2019, Rizzo2023, RomanOliveira2023}, pressure support is increasingly important and the most likely case is a combination of these two effects. 

We fit the major-axis rotation curve with three different models (see Table~(\ref{tab:model components})), including a NFW halo, ring/disk and a central bulge. We distinguish between the case (A) of a massive ``dynamical ring" and (B) an exponential disk with a ``flux emitting ring". We include a case (C) where a contracted DM halo might replace a central stellar bulge. We find that our model (A) gives the best fit. The derived ring mass is$M_{\mathrm{ring}} = 1.2-2.5 \times 10^{10}\, \mathrm{M_\odot}$ with a peak radius of $R_{\mathrm{p}} = 4.6\, \mathrm{kpc}$ and a shape parameter of $h = R_{\mathrm{p}}/\mathrm{FWHM_{ring}} = 0.89$. Our results indicate the presence of a massive bulge $M_{\mathrm{bulge}} = 8 \times 10^{10}\, \mathrm{M_\odot}$ and a highly baryon-dominated system with $f_{\mathrm{DM}}(R_{\mathrm{eff}}) = 0.07$, fully consistent with previous modeling that did not explicitly consider a ring component. 

Using archival ALMA observations we constrain the molecular gas and dust mass. As these observations are poorly resolved, it might be distributed in such a way as to attenuate the light from the central bulge, explaining the lack of any detection. We show that, using standard assumptions for the M/L of the bulge and the attenuation magnitude, this mass is more than enough to fully attenuate the rest-frame g-, i- and J-bands (\emph{JWST}/NIRCam F150W, F277W and F444W, respectively). However, it requires that $\sim 10\%$ of the dust acts as a screen between the bulge and the observer. This scenario can be directly tested, as the continuum emission from the attenuating dust should be observed with \emph{ALMA} at band 4, with a total time of around 10 hours at $0.25^{\prime \prime}$ resolution.

A CO(4-3) flux and/or dust continuum detection at the center from sufficiently high resolution and sensitivity ALMA data, as well as a spectroscopic redshift for the western neighbor, are fundamental in order to elucidate the nature of this intriguing ring galaxy and shed light on important processes in the evolution of galaxies. \\

\textit{acknowledgments} This work is based in part on observations made with the NASA/ESA/CSA James Webb Space Telescope. The data were obtained from the Mikulski Archive for Space Telescopes at the Space Telescope Science Institute, which is operated by the Association of Universities for Research in Astronomy, Inc., under NASA contract NAS 5-03127 for JWST. The {\it JWST} data used in this paper are associated with program \#1727 and can be found in MAST: \dataset[10.17909/x7ct-ef43]{http://dx.doi.org/10.17909/x7ct-ef43}. This paper makes use of the following ALMA data: ADS/JAO.ALMA\#2013.1.00668.S, ADS/JAO.ALMA\#2017.1.01020.S. ALMA is a partnership of ESO (representing its member states), NSF (USA) and NINS (Japan), together with NRC (Canada), NSTC and ASIAA (Taiwan), and KASI (Republic of Korea), in cooperation with the Republic of Chile. The Joint ALMA Observatory is operated by ESO, AUI/NRAO and NAOJ.
This work was supported by the German Science Foundation via DFG/DIP grant STE/1869-2 GE 625/17-1. ANS and AS are supported by the Center for Computational Astrophysics (CCA) of the Flatiron Institute, and the Mathematics and Science Division of the Simons Foundation, USA. H{\"U} acknowledges funding by the European Union (ERC APEX, 101164796). TN acknowledges the support of the Deutsche Forschungsgemeinschaft (DFG, German Research Foundation) under Germany’s Excellence Strategy - EXC-2094 - 390783311 of the DFG Cluster of Excellence ``ORIGINS". NMFS, CB, JC, JMES and GT acknowledge funding by the European Union (ERC Advanced Grant GALPHYS, 101055023). Views and opinions expressed are, however, those of the author(s) only and do not necessarily reflect those of the European Union or the European Research Council. Neither the European Union nor the granting authority can be held responsible for them.

\begin{software}
We have made use of \texttt{astropy} \citep{AstropyCollaboration2013}, \texttt{scipy} \citep{scipy2020}, \texttt{numpy} \cite{numpy2011}, \texttt{matplotlib} \citep{matplotlib2007}, \texttt{emcee} \citep{Foreman-Mackey2013}, \texttt{pandas} \citep{pandas2022}, \texttt{Bagpipes} \citep{Carnall2018_bagpipes} and \texttt{dysmalpy} \citep{Davies2004_dysmalpy, Davies2004b_dysmalpy, Cresci2009_dysmalpy, Davies2011, Wuyts2016, Lang2017, Price2021_rc41, Lee2025}.
\end{software}

\bibliographystyle{mnras}
\bibliography{references.bib}{}

\appendix

\section{SED fitting of the western companion}\label{app:SED_fitting}
We extracted the integrated total fluxes within a circular aperture of 0.5$^{''}$ radius centered on the companion, for the total set of observed bands (HST/WFC3: F438W, F814W, F110W, F160W; JWST/NIRCam: F115W, F150W, F277W, F444W). We fit the SED using the code \textsc{bagpipes} \citep{Carnall2018}, assuming a \cite{Kroupa2001} initial mass function, a \cite{Calzetti2000} attenuation curve, and \cite{BruzualCharlot2003} stellar population models. We fix the metallicity to a fiducial Solar value and adopt an exponentially declining star formation history. The visual extinction is bound between $0<A_{\mathrm{V}}<5$ and the redshift $0<z_\mathrm{phot}<4$. The median $z_\mathrm{phot} = 2.14^{+0.32}_{-0.42}$ (16$^{th}$ and 84$^{th}$ percentiles), within the redshift range of zC406690, but the probability distribution is very broad. The estimated stellar mass is of course degenerate with $z_\mathrm{phot}$ and is $\log \left( M_\star / M_\odot \right) =10.25^{+0.16}_{-0.17}$, giving a mass ratio of 2.25:1 compared to the stellar mass of zC406690. 
\begin{figure}[ht]
    \centering
    \includegraphics[width=\columnwidth]{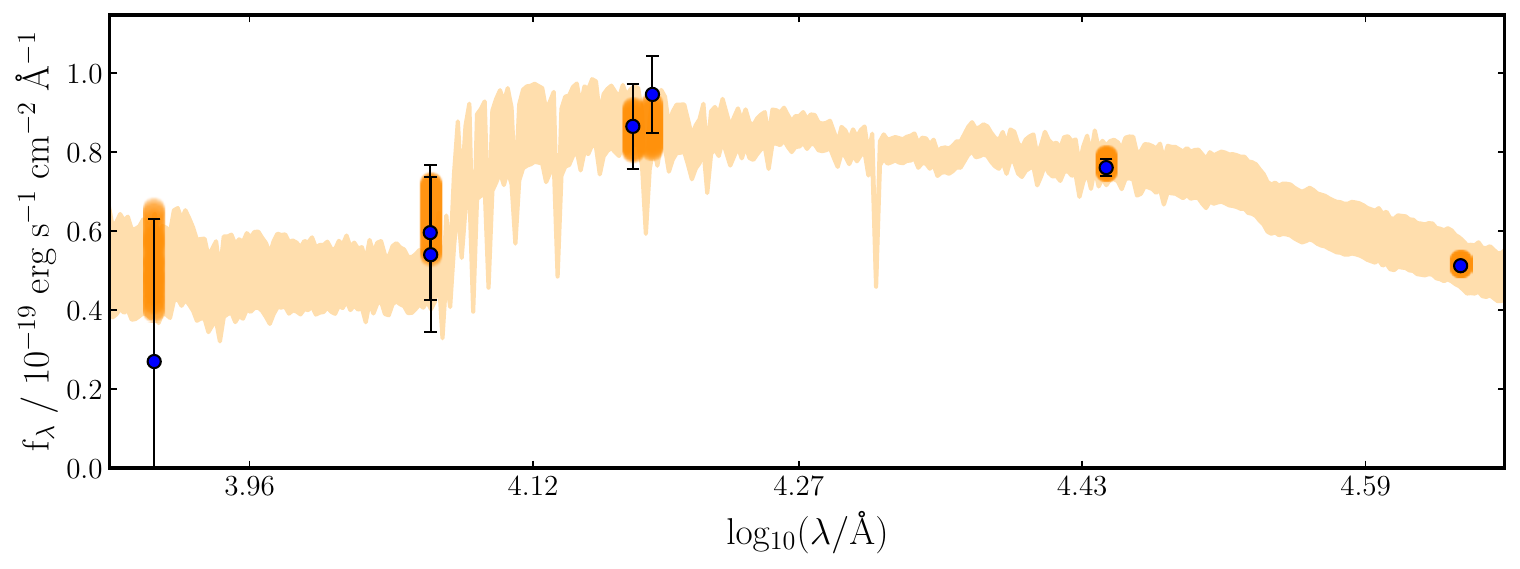}
    \caption{Spectral Energy Distribution (SED) for the western companion of zC406690. The observations are shown in black, and the best fit ($\pm 1 \sigma$) given by the shaded region.}
    \label{fig:SED_fitting_fit}
\end{figure}

\begin{figure}[ht]
    \centering
    \includegraphics[width=\columnwidth]{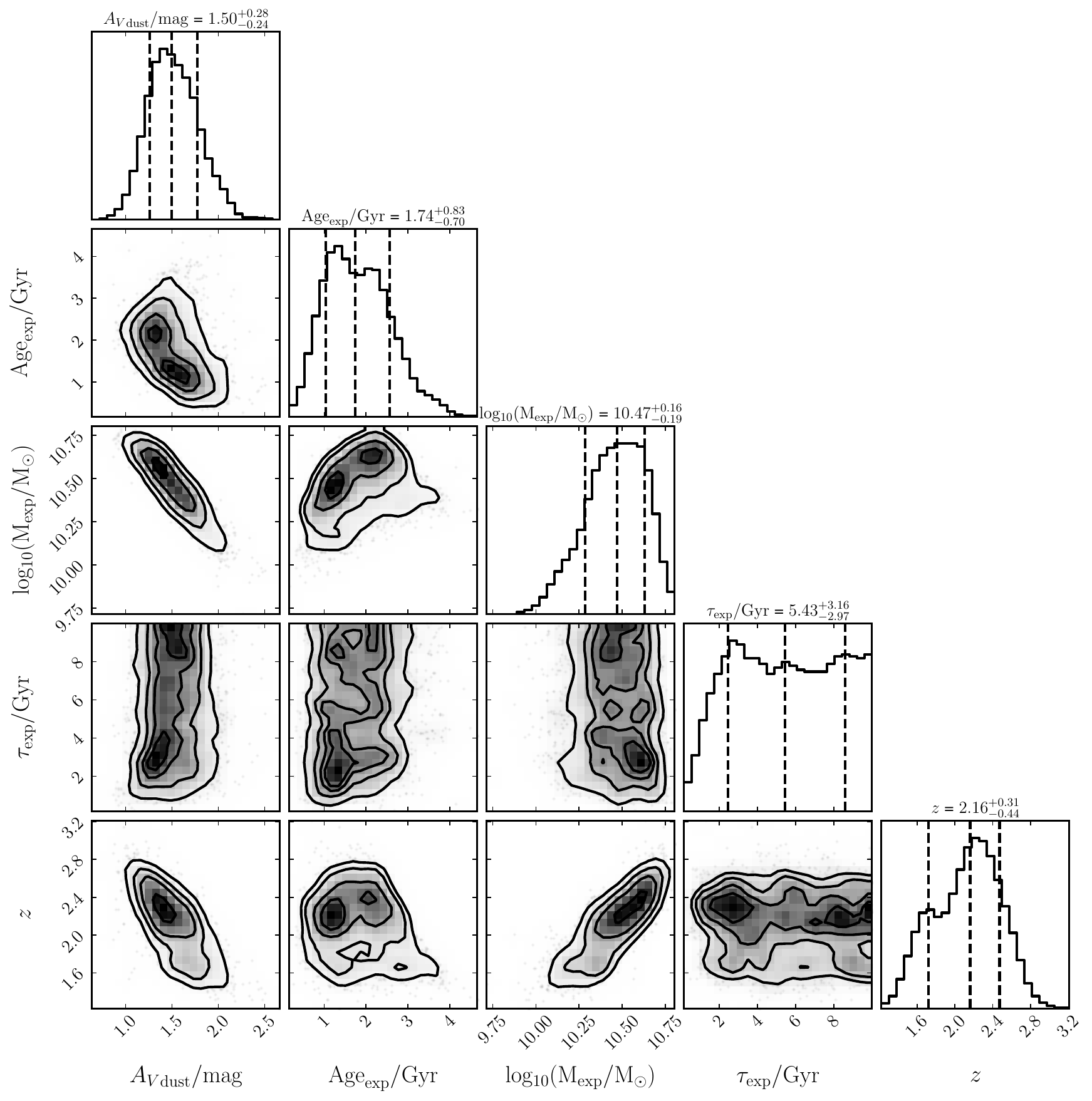}
    \caption{Corner plot for the SED fitting of the companion.}
    \label{fig:SED_fitting_corner}
\end{figure}

\section{Gaussian Ring mass distribution}\label{app:GaussianRing}
\subsection{General expressions}\label{app:GaussianRing_general_equations}
\begin{figure}[ht]
    \centering
    \includegraphics[width=\columnwidth]{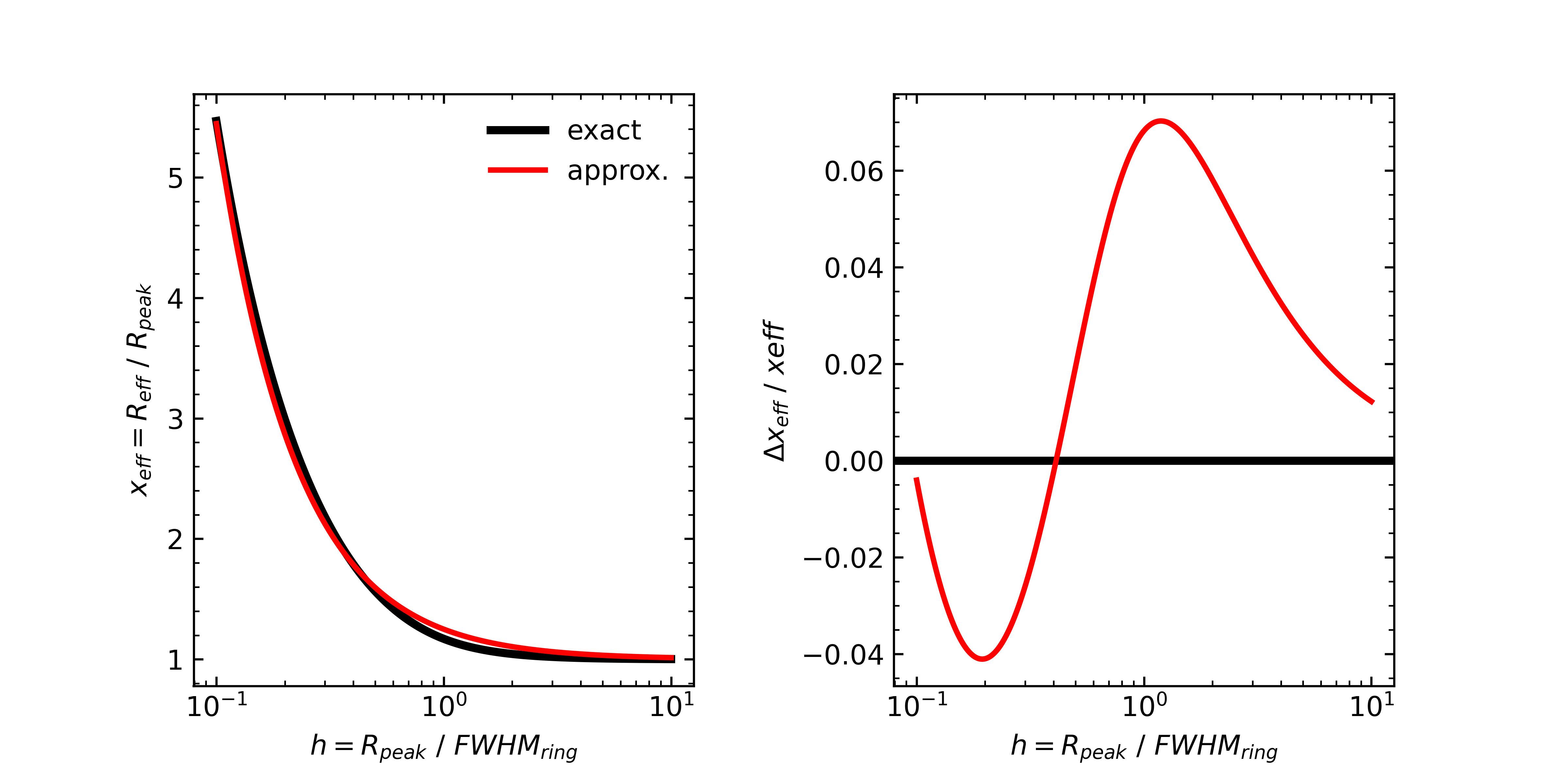}
    \caption{Effective (half mass) radius of a Gaussian surface density profile as a function of the shape parameter $h$. Left: $x_{\mathrm{eff}}$ as a function $h$ (black) and the approximation (red) given by $x_{\mathrm{eff}} = 1 + \frac{1}{4} h^{-5/4}$. Right: Difference between the approximation (red) to the real value.}
    \label{fig:Gaussian_ring_xeff}
\end{figure}

We define a razor-thin, axisymmetric mass distribution with a total mass $M_{\mathrm{ring}}$ defined by three parameters: Peak radius $R_\mathrm{p}$, the standard deviation $\sigma_{\mathrm{r}}$ (or the FWHM $F_r = \sqrt{8\, \ln{2}} \, \sigma_{r}$) and a characteristic surface density $\Sigma_0$:
\begin{equation} \label{eq_app:gaussian_profile}
    \Sigma(r) = \Sigma_0 \times exp\left\{-\frac{(r-R_p)^2}{2\sigma_{r}^2}\right\}
\end{equation}
Next, we define a dimensionless radius $x=r/R_\mathrm{p}$ and a dimensionless ``shape" parameter $h \equiv R_{\mathrm{p}} / F_{\mathrm{r}}$. The shape parameter is the only parameter determining the behavior of this function, as it encodes the normalized width of the ring. If the peak radius is larger than the FWHM the ring will have a larger ``hole" in its center, and vice versa. We refer to rings with values of $h \gg 1$ as ``narrow" and rings with $h \ll 1$ as ``wide" (see Figure~\ref{fig:GaussianRing_dimless_densityMass}). 
The dimensionless form of Equation~(\ref{eq_app:gaussian_profile}) is:
\begin{equation}\label{eq:dimless_gaussian_ring_surface_density}
\begin{split}
    \Sigma(x) &= \Sigma_0 \times f_\Sigma(x) \\
    f_\Sigma(x) &= \exp {\left\{- 4\, \ln{2}\, h^2 (x-1)^2 \right\}} \ \ \ .
\end{split}
\end{equation}
The enclosed mass within radius $x$ is can be expressed in terms of the total mass in the ring:
\begin{equation}
    M (<x) = M_{\mathrm{ring}} \frac{1}{f_{\mathrm{M}}} \int_0^{x} {t \, f_{\Sigma}(t) dt}
\end{equation}
where we have used the total mass of the ring:
\begin{equation}
\begin{split}    
    M_{\mathrm{ring}} &= 2 \pi R_{\mathrm{p}}^2 \Sigma_0 \times f_{\mathrm{M}} \\
    f_{\mathrm{M}} & = \int_0^{\infty} {t \, f_{\Sigma}(t) dt} \approx \left( h - 0.07 \right) ^{-1} \ \ \ .
\end{split}
\end{equation}
The approximation in the second row is valid for $h > 0.15$. 
The effective radius, the radius enclosing half of the total mass, is given by requiring $M( < x_{\mathrm{eff}}) = \frac{1}{2} M_{\mathrm{ring}}$. Thus, the value of $x_{\mathrm{eff}}$ depends only on the shape parameter $h$. From geometrical considerations we expect to have $x_{\mathrm{eff}} > 1$, as more mass will reside beyond the peak radius than inside it. For an infinitely narrow ring, we must have $x_{\mathrm{eff}} = 1$ as all of the mass is concentrated on $r=R_{\mathrm{peak}}$. Figure~\ref{fig:Gaussian_ring_xeff}) shows how $x_{\mathrm{eff}}$ varies with the shape parameter, and that it can be well approximated by:
\begin{equation}\label{eq:GaussRing_xeff}
x_{\mathrm{eff}} \approx 1 + \frac{1}{4} h^{-5/4}
\end{equation}
for values of $0.1 < h < 10$. 

The gravitational potential along the mid-plane has no analytical formula, but it can be expressed in terms of the general potential for a razor-thin mass distribution \citep[following][eq. 2.155]{Binney1987}. Inserting Equation~(\ref{eq:dimless_gaussian_ring_surface_density}), the gravitational potential $\phi_{\mathrm{ring}}$ is given by: \\
\begin{equation}\label{eq:dimless_Gaussian_potential_velocity}
\begin{split}
    \phi_{\mathrm{ring}}(x) & = \frac{G M_{\mathrm{ring}}}{R_\mathrm{p}} \times f_\phi(x) \\
    f_{\phi} (x) &= \frac{1}{f_{\mathrm{M}}(h)}\, \int_0^x {{\frac{- I(t)}{\sqrt{x^2-t^2}} dt}} \equiv \frac{G M_{\mathrm{ring}}}{R_{\mathrm{p}}}
\end{split}
\end{equation}
where we have defined the intermediate function:
\begin{equation}\label{eq:I_definition}
    I(t) = \int_{0}^{\infty} {f_{\Sigma} \left( \sqrt{s^2 + t^2} \right) ds}
\end{equation}
the circular velocity, $V^2_{\mathrm{ring}} = r \frac{d\phi_{\mathrm{ring}}}{dr}$, is:
\begin{equation}\label{eq_app:circ_velocity}
\begin{split}
    V_{\mathrm{ring}}^2 (x) & = \frac{G M_{\mathrm{ring}}}{R_{\mathrm{p}}} \times f_{\mathrm{V}} (x) \\
    f_{\mathrm{V}} (x) &=  \frac{1}{f_{\mathrm{M}}(h)} \, \int_0^x { \frac{- dI(t)}{dt} \frac{t}{\sqrt{x^2 - t^2}} dt } \equiv \frac{G M_{\mathrm{ring}}}{R_{\mathrm{p}}} 
\end{split}
\end{equation}
with $\frac{dI}{dt}$ given by differentiating Equation~(\ref{eq:I_definition}):
\begin{equation}
\begin{split}
     \frac{dI(x)}{dt} & = \int_{0}^{\infty} {\frac{d f_{\Sigma} (s) }{ds} \frac{t}{\sqrt{s^2 - t^2}} ds} \\
\end{split}
\end{equation}
for the density profile defined in Equation (\ref{eq_app:circ_velocity}) these terms cannot be solved analytically, but they can be evaluated numerically. However, these expressions are uniquely determined by a single parameter, the shape parameter $h$, and are only scaled by the total mass and peak radius. Figures~(\ref{fig:GaussianRing_dimless_densityMass}) and~(\ref{fig:GaussianRing_dimless_potentialVelocity}) show the behavior of these functions for values of $h=0.5, 1, 2$. The gravitational potential in the central regions has a clear negative gradient, meaning a centrifugal equilibrium cannot be supported by self-gravity alone, and a stabilizing component has to be included in order for the system to be rotationally stable.

\subsection{Rotational equilibrium}\label{app:GaussianRing_circularEquilibrium}
\begin{figure}[ht]
    \centering
    \includegraphics[width=\columnwidth]{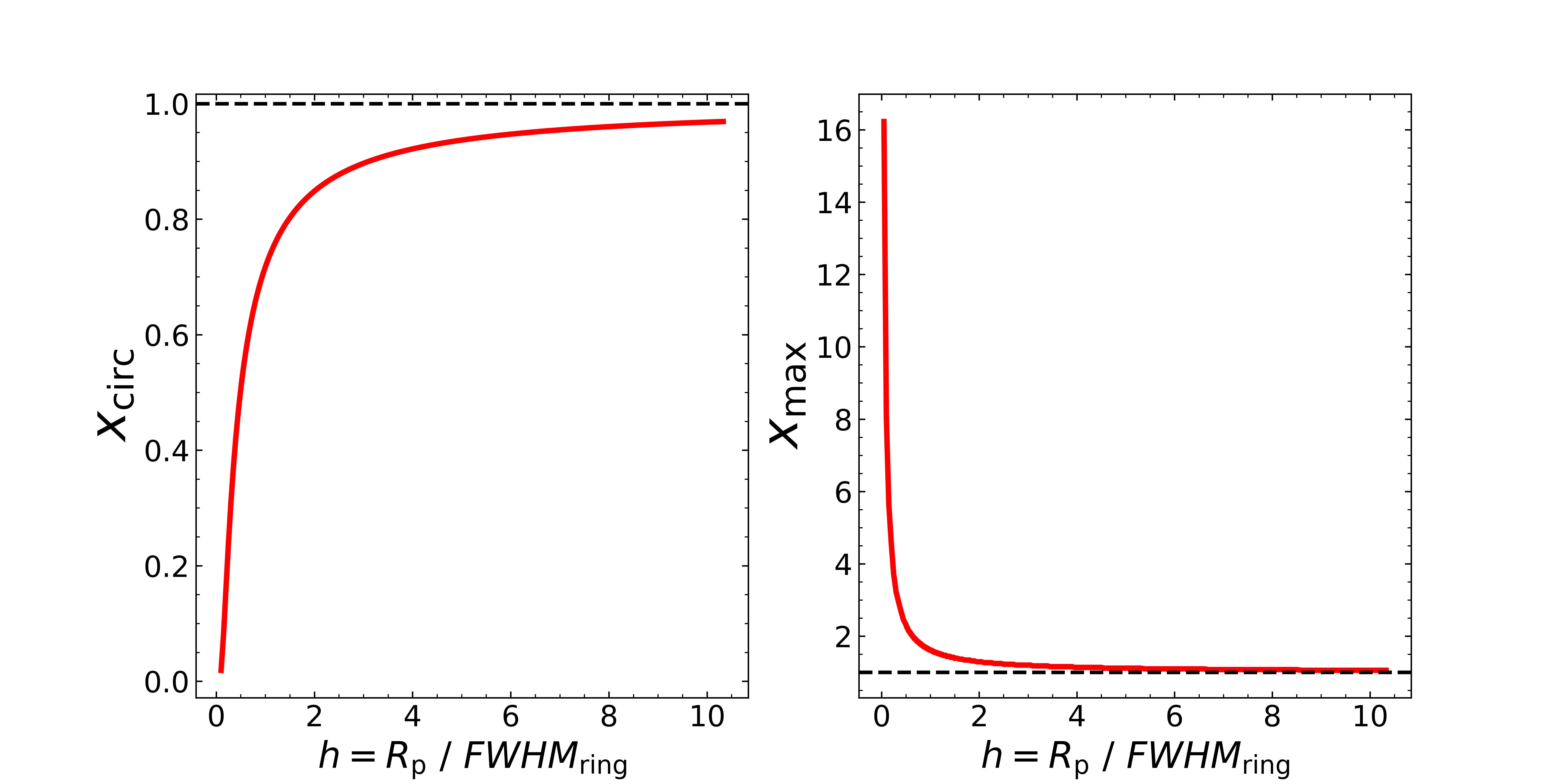}
    \caption{Left: Circular equilibrium radius for a Gaussian ring as a function of the shape parameter. For $x < x_{\mathrm{circ}}$ there are no circular orbits sustaining a centrifugal equilibrium with the ring gravitation potential. Right: The radius $x_{\mathrm{max}}$ where the circular velocity peaks at.}
    \label{fig:Gaussian_ring_xstability}
\end{figure}

\begin{figure}[ht]
    \centering
    \includegraphics[width=\columnwidth]{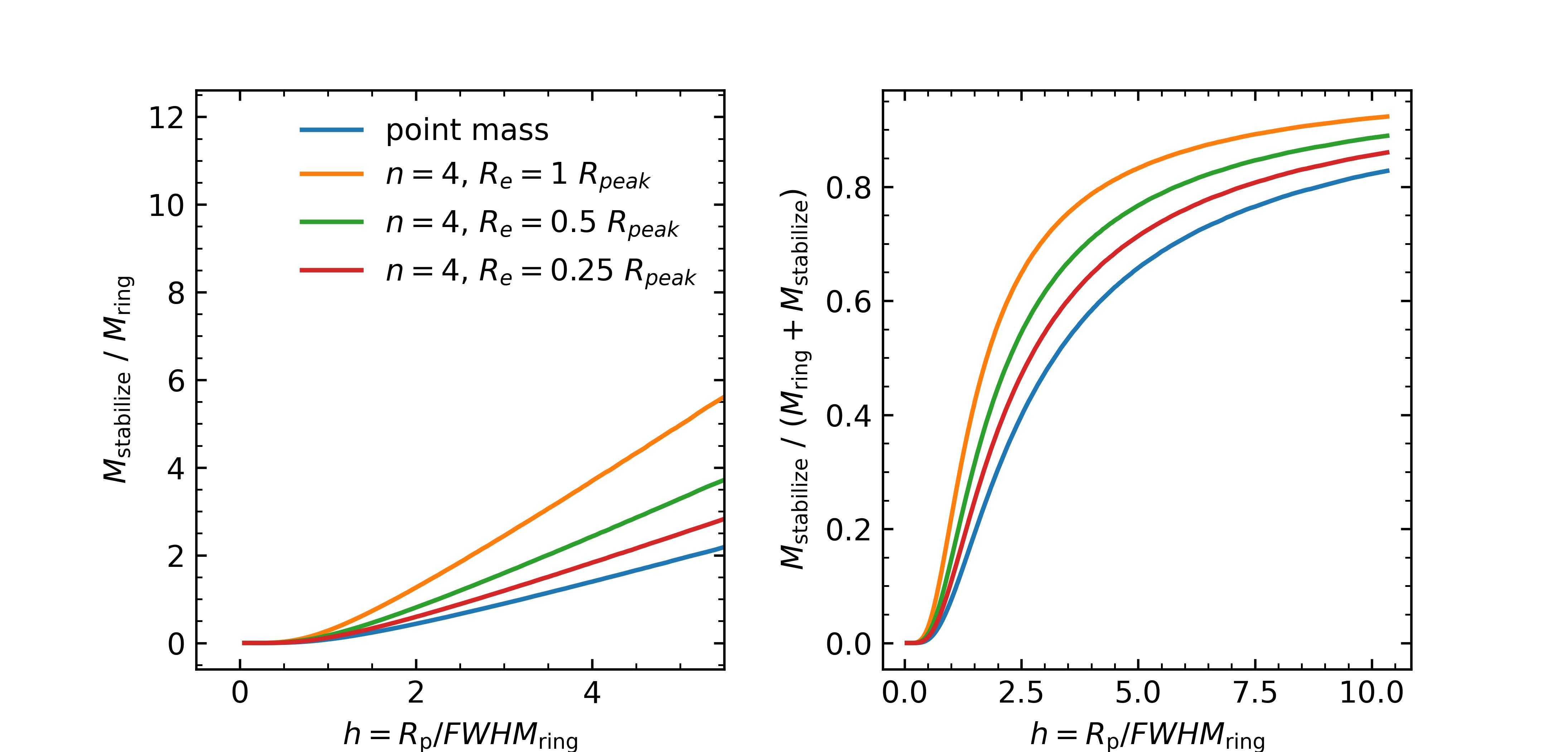}
    \caption{The minimal total mass $M_{\mathrm{stabilize}}$ of the second component required for the ring to be in centrifugal equilibrium, i.e., satisfy Equation~(\ref{eq:Gaussian_ring_stabilizing_criteria}). Different colors refer to a point mass, or Sérsic $n_s=4$ profiles with various sizes relative to the ring, representing a central bulge.}
    \label{fig:stabilizing mass}
\end{figure}

The circular velocity term $V_{\mathrm{ring}}^2$ given by Eq.~(\ref{eq:dimless_Gaussian_potential_velocity}) is not properly defined at all radii, as an outwards net force is applied to particles in the inner regions of the Gaussian Ring. Mathematically, this is because the surface density is not decreasing everywhere, but increasing up to the peak radius $R_p$,  creating regions where $\frac{dI(t)}{dt} > 0$ leading to a negative gradient in the potential. The narrower ($ h \gg 1$) the ring is, the larger this region becomes until it covers the entire area within $R_p$. 
We define the circular radius, $x_{\mathrm{circ}}$, as the radius from which a test particle can sustain rotation equilibrium with the ring's self-gravity alone. Figure~\ref{fig:Gaussian_ring_xstability} shows that narrower rings (large $h$) become stable to rotation at increasingly larger radii, approaching $R_{\mathrm{p}}$. For our fiducial $h=1$ we get $x_{\mathrm{circ}} = 0.7$.
It is clear that any mass distribution following the Gaussian ring density profile cannot be rotationally supported on its own. To account for it, keeping with the mass model we use in this paper, we introduce a second compact mass component and require it to be massive enough to sufficiently deepen the gravitational potential. We assume a second mass distribution with a total mass $M_{\mathrm{bulge}}$ and a well-defined circular velocity $V^2_{\mathrm{bulge}} (r)$. In order to have a rotationally stable system, we look for the minimal mass for which the total circular velocity will be defined at all radii:
\begin{equation} \label{eq:Gaussian_ring_stabilizing_criteria}
    V_{ring}^2(x) + V_{\mathrm{bulge}}^2(x) > 0
\end{equation}
since $x_{\mathrm{circ}} < 1$, this condition is only needed to be checked for this radial range.
If the stabilizing mass distribution is spherically symmetric, Equation~(\ref{eq:Gaussian_ring_stabilizing_criteria}) can be combined with Equation~(\ref{eq_app:circ_velocity}) to become:
\begin{equation}\label{eq:Gaussian_ring_stabilizing_criteria_circ_mass}
    \frac{G\, M_{\mathrm{ring}}}{R_p\, x} \left[ x\, f_{V}(x) + \frac{M_{\mathrm{bulge}}(<x)}{M_{\mathrm{ring}}} \right] > 0
\end{equation}
and since $f_{V} (x)$ is a function of $h$, we can find the mass ratios required as a function of the ring shape. In figure~(\ref{fig:stabilizing mass}) we consider two cases: (i) an idealized point mass, i.e., $M_{\mathrm{bulge}} (<x) = M_{\mathrm{bulge}}$, the simplest case serving as a lower-limit; (ii) a physically motivated spherically symmetric Sérsic profile ($n_{\mathrm{s}} = 4$) with various ring-to-bulge effective radii ratios (all smaller than the ring peak radius). In general, narrower rings need more massive counterparts and the more extended the component considered the more massive it needs to be. For fiducial values of $h = 1$ and $R_{\mathrm{eff}} = 0.25 R_p$ the central bulge needs to be only at least $15 \%$ of the ring's mass in order for the system to be rotationally stable.

\subsection{Impact of a Flux Ring }\label{app:GaussianRing_fluxRing}
\begin{figure*}[t]
    \centering
    \includegraphics[width=18cm]{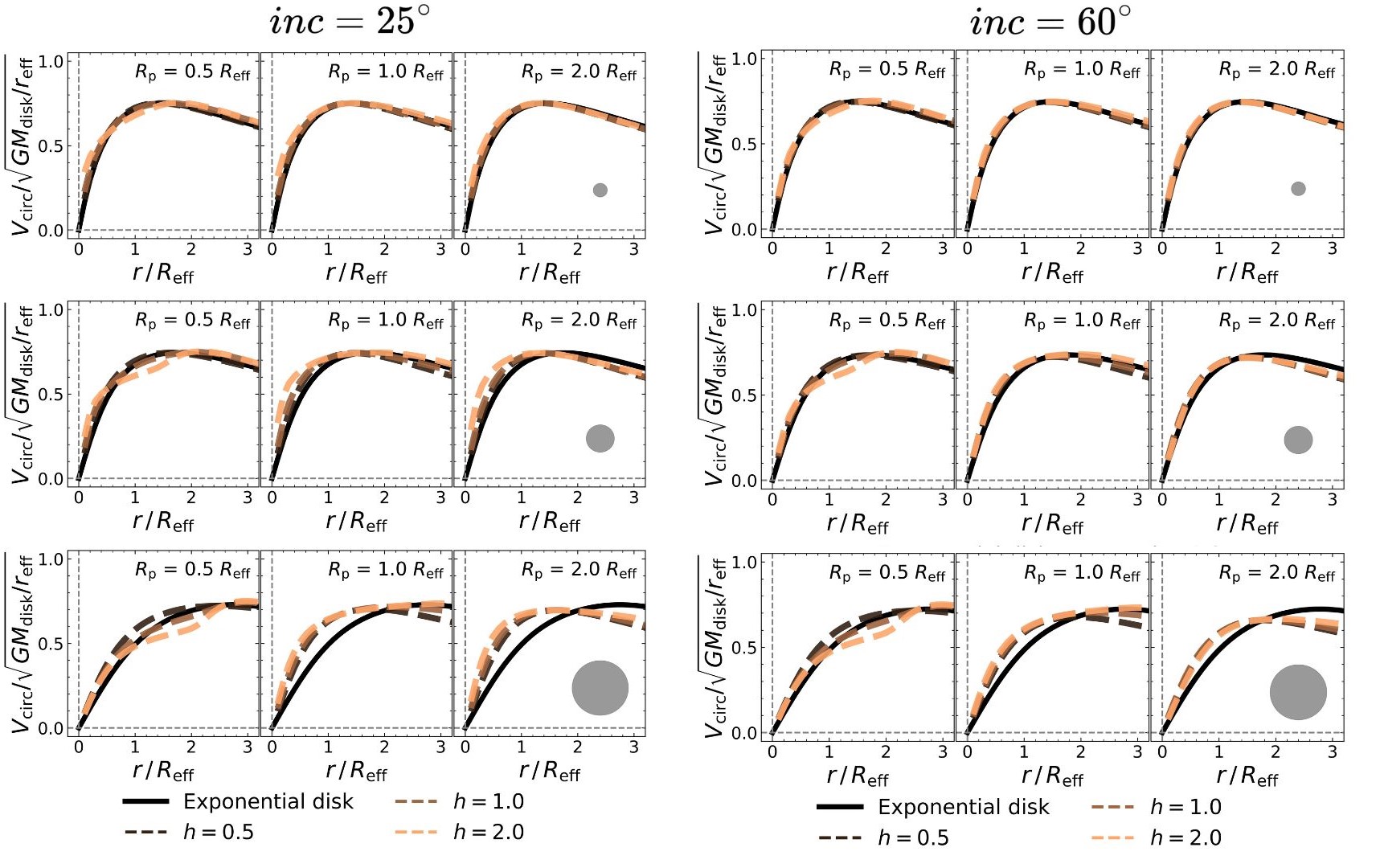}
    \caption{The effect of a ``flux ring” on the observed circular velocity for a pure exponential disk (similar to Figure~\ref{fig:flux_ring_effects}) for different inclinations and beam PSF FWHM sizes. The left column has an inclination of inc$=25^{\mathrm{o}}$ and the right column inc$=60^{\mathrm{o}}$. Each row has an increasing beam size with respect to the radius of the flux emitting ring, ranging from a PSF FWHM of $(0.25, 0.5, 1.0) \times R_{\rm p}$ for the top, middle and bottom row respectively.}
\label{fig:app_flux_ring_effects}
\end{figure*}

Resolution effects (i.e., ``beam smearing") are affected by the flux distribution of the observed tracer, and in general are very sensitive to the size of the beam PSF (with respect to the size of the system) and the inclination angle. In \S~\ref{sec:3.3light_weighting} we examine the effects of a (dark) exponential disk with a flux emitting ring and examine the implications on the observed RC. As this effect tends to increase as the inclination is more face-on and as the PSF size increases, Figure~(\ref{fig:app_flux_ring_effects}) examines a few more cases with different choices of the inclination angle and PSF FWHM.

\newpage

\section{Rotation Curve fitting with \texttt{dysmalpy}}\label{app:RC_fitting_dysmalpy}
\begin{figure*}[t]
    \centering
    \includegraphics[width=0.65 \linewidth]{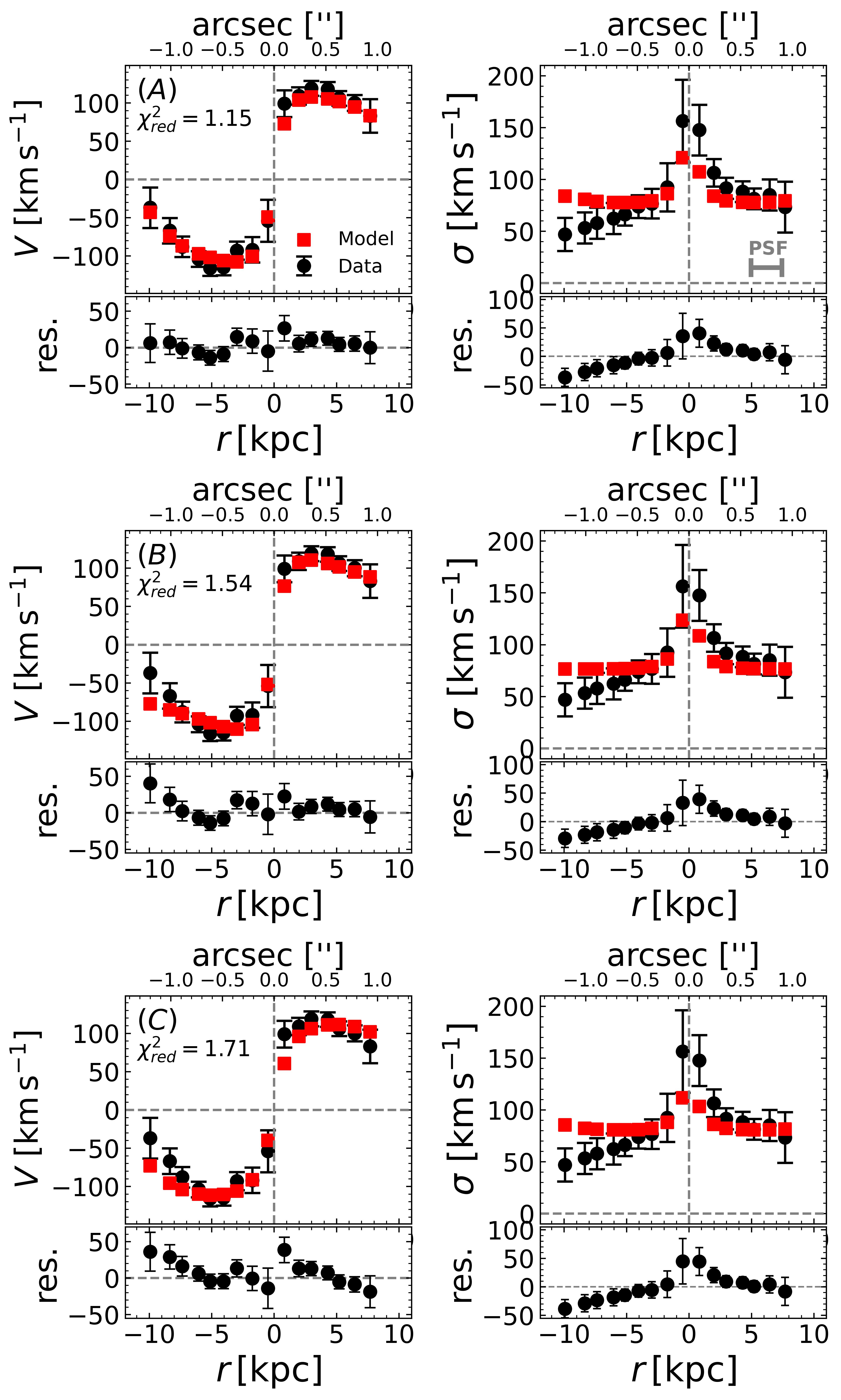}
    \caption{Similar to Figure~(\ref{fig:RC_bestfit}), but using \texttt{dysmalpy}.}
    \label{fig:RC_bestfit_dysmalpy}
\end{figure*}

As described in \S~\ref{sec:4.2RC fitting}, we fit the RC using the code \texttt{dysmalpy} for the three mass models considered in Table~(\ref{tab:model components}). The best fit result using \texttt{dysmalpy} are given in Figure~(\ref{fig:RC_bestfit_dysmalpy}). We find that all best-fitting values agree well within their uncertainties for each of the three models respectively, when comparing both fitting procedures.
\end{document}